A

# Crowdsourcing through Cognitive Opportunistic Networks

M. Mordacchini, A. Passarella, M. Conti, Istituto di Informatica e Telematica, CNR, Italy.
S.M. Allen, M.J. Chorley, G.B. Colombo, V. Tanasescu and R.M. Whitaker, School of Computer Science & Informatics, Cardiff University, UK

Until recently crowdsourcing has been primarily conceived as an online activity to harness resources for problem solving. However the emergence of opportunistic networking (ON) has opened up crowdsourcing to the spatial domain. In this paper we bring the ON model for potential crowdsourcing in the smart city environment. We introduce cognitive features to the ON that allow users' mobile devices to become aware of the surrounding physical environment. Specifically, we exploit cognitive psychology studies on dynamic memory structures and cognitive heuristics, i.e. mental models that describe how the human brain handles decision-making amongst complex and real-time stimuli. Combined with ON, these cognitive features allow devices to act as proxies in the cyber-world of their users and exchange knowledge to deliver awareness of places in an urban environment. This is done through tags associated with locations. They represent features that are perceived by humans about a place. We consider the extent to which this knowledge becomes available to participants, using interactions with locations and other nodes. This is assessed taking into account a wide range of cognitive parameters. Outcomes are important because this functionality could support a new type of recommendation system that is independent of the traditional forms of networking.

## 1. INTRODUCTION

Through the advent of the smartphone and other enabling wireless technologies there are now increased interaction between mobile devices, the physical environment and data sources within it. This scenario is known as the converging *Cyber-Physical World* (CPW) [Conti et al. 2012] and within this, opportunistic networking [Boldrini and Passarella 2013] is an important enabling paradigm. Opportunistic networking is an entirely self organised form of communication which functions by mobile devices, such as smartphones, transiently connecting when they come into range. This is known as the store-carry and forward paradigm, which opens up new ways to create and share knowledge through data dissemination. This makes them ideal for crowdsourcing applications where distributed resources are harnessed to provide new services for a wide range of emerging applications including smart cities, e-health, intelligent transportation systems [Conti et al. 2012]. Unlike many forms of online crowdsourcing, opportunistic networking differs in that the providers of resources are also the consumers of sub-services, described by the subset of data that is relevant for their needs and interests. This participatory prosumer model is a distinctive feature of opportunistic networking.

The concept of opportunistic networking brings networking closer to the disposition of the human user because the mobile devices such as smartphones which perform the networking function are carried around throughout the users day-to-day activity. Due to this these devices can act as cyber-physical *proxies* for their human users [Whitaker et al. 2015], potentially autonomously discovering, collecting and evaluating data from opportunistic collection, determining its relevance and making sense of it in relation to other data. Due to this role of proxies, it is interesting to investigate if algorithms for communication and data exchange in opportunistic networks can follow rules and behaviours similar to those governing the functions of the human brain. This approach exploits a much stronger link between users and devices with respect to conventional bio-inspired solutions. The motivation of this approach is to make the devices' behaviour as similar as possible to that their human users would have if they had to process the same information coming from the environment around them, gathered from both physical artefacts and other users they physically meet. Indeed, humans are continuously presented with a vast amount of stimuli from the physical environment within which they are immersed. It is a distinctive feature of human intelligence that the brain is able to swiftly process and react to information, making decisions under highly constrained conditions with little resources, to the extent that many processes occur subconsciously. When combined with effective memory structures, it has been hypothesised that this capability is enabled by the use of simple cognitive decision-making rules, known as *cognitive heuristics* [Gigerenzer 2008]. These represent "computationally inexpensive" rules, that are functionally described in the cognitive science literature. These are potentially highly valuable for opportunistic networking scenarios where protocols need to make rapid decisions under constrained conditions.

In this paper we propose a solution that allows each device to learn and keep in memory a global representation of the semantic pieces of data (given in terms of *keywords* or *tags*) that describe the surrounding environment, the relationships between them (i.e. how they could co-occur in describing specific places), and how they are associated with physical locations. This solution is based on an opportunistic exchange of information between devices and physical locations. Specifically, in this paper we enhance crowdsourcing in opportunistic networking by applying features of cognition and memory. This allows us to structure and interact with heterogeneous data for intelligent dissemination. The scenario that we address is motivated by mobile devices acting as proxies for their human owners, where mobile devices become aware of features about places in the physical environment through which they move. The opportunistic interactions between devices occur in two ways: firstly through mobile devices interacting with static devices, such as those that may be found at fixed locations and places; secondly through mobile to mobile device interactions. The challenge is to efficiently disseminate the heterogeneous information about the physical places so that it is received by users who will find it relevant. Efficiently here means that information should be brought ideally to all interested users, and only to them, by using resources available at nodes in the best way. The goal of this paper is showing that by using cognitive heuristics we can build crowd sourcing systems based on opportunistic networking that are more efficient than state of the art solutions that do not exploit cognitive approaches.

The heterogeneous data that we consider are *tags*, being single word semantic descriptions concerning the characteristics of places. We develop memory structures, inspired by *semantic associative network models* of human memory that allow individual devices to organise and prioritise interaction with each other for exchange of information. This is driven by models of how humans exchange information during a discussion where common concepts are often a starting point for conversation and exploration of further associated concepts from individual perspectives. We capture this by exploiting a *fluency heuristic* [Schooler et al. 2005] strategy. This is a cognitive heuristic from psychological origins that captures how the brain may rapidly choose among two or more alternatives, with preference given to those that have been previously recognised before in some form.

We show how this approach to cognitive opportunistic networks provides an efficient new form of relevant dissemination and we establish a comparison with alternative standard solutions that do not adopt such heuristics. We also find that the semantic memory networks at individual nodes are structurally very similar to the asymptotic one, containing the union of all physical locations' descriptions, that would be collected with infinite available resources. We also analyse the behaviour of the proposed solution when locations are allowed to adapt the relevance of the tags in their descriptions in response to the interactions with the devices of users that visit them. Moreover, we study the performance of the system when users and locations are divided in different communities, place in geographically separated area and only a restricted set of devices allow the flow of information between different communities.

## 2. RELATED WORK

Data dissemination in opportunistic networks is a fundamental problem [Boldrini and Passarella 2013] that underpins future applications of the technology. The hidden structures in human mobility [Williams et al. 2012] offer a way in which this can be harnessed for new candidate applications [Allen et al. 2010; Allen et al. 2012]. Due to these features data dissemination has received a lot of attention in the field. However to date, the approaches taken are based on computer science based solutions that are generally constructive, probabilistic or structural, often based around inspiration from networks. Only relatively recent work [Conti et al. 2011; Bruno et al. 2012; Valerio et al. 2013] has started to consider cognitive science as an alternative fields of inspiration to devise simple and low resource-demanding schemes for efficient and effective dissemination. The rationale for this approach comes from the observation that mobile devices are increasingly acting as proxies for humans and facing similar issues concerning the volume, velocity and variety of stimuli. Therefore it is possible that cognitive heuristics used by the brain may prove effective, particularly as it is very effective in dealing with such problems with low resource overhead [Conti et al. 2013a; Mordacchini et al. 2014].

In [Conti et al. 2013a; Bruno et al. 2012; Valerio et al. 2015] the cognitive based solutions proved to be as efficient as other alternative "traditional" approaches in disseminating the data toward interested nodes, while, at the same time, requiring much lower resources to achieve the same performance. These works were a useful proof of concept but were limited to homogeneous data with no consideration of different data types and their association. To address this, in [Conti et al. 2013b] an algorithm for spreading semantic information has been proposed. Beyond this the proof of concept in [Mordacchini et al. 2013b] represents, to the best of our knowledge, the first attempt to equip nodes with cognitive-based capabilities to exchange and diffusion of semantic information and its associated content. All these works do not consider possible interactions between mobile devices and the physical environment the nodes are moving in. Moreover, also the data available in the network is assumed to be not related in any way to the physical context. Thus, by using those schemes, it is not possible for nodes in an opportunistic network to become aware of their surrounding environment.

Other approaches make use of nature-inspired mechanisms to let mobile devices become aware of features in their physical surroundings. For instance, systems based on the dissemination of digital pheromones [Mamei and Zambonelli 2007] allow the devices to localise and track other physical objects moving in the same space, using interactions with RFID tags dispersed in the environment. On the other hand, using computational fields [Mamei and Zambonelli 2009], information about objects and other devices is presented in form of tuples that are spread in the environment and allow localisation of interesting items or to put in place spatial coordination behaviours among nodes to achieve specific goals.

Rather than the acquisition of information about single, distinct items, in this paper we propose a solution that let nodes to gather and construct a global vision of the semantic representation of information about physical locations. To reach this goal, we show how cognitive memory representation and recall (i.e., information selection schemes) taken from cognitive psychology can be exploited. A preliminary analysis of such a solution is given in [Mordacchini et al. 2013a]. In this paper we extend that work by exploring more dynamic and complex scenarios, where information about physical locations change over time, and users move according to different social groupings.

In order to communicate their features to mobile nodes passing near-by, physical locations need to generate their own descriptions. Virtual representations of place have been the topic of active research, with an understanding that, to the human mind, the geospatial environment is organised as places rather than sets of geospatial coordinates [Egenhofer and Mark 1995]. Places appear as complex thematic entities in relation with the physical configuration of the environment as well as with human cognition such as memory [Tuan 2001]. Places, as part of the environment, present physical or social opportunities for action or information [Raubal et al. 2004], as well as for social interaction with other agents, or the absence of them [Clark and Uzzell 2002]. Therefore, they are spatial regions that support information of significance for the agent in an environment and act as cognitive anchors, through salient features that make them useful or interesting, memorable because of past experiences, or desirable as expected loci of anticipated ones [Gibson 1986; Jordan et al. 1998; Bennett and Agarwal 2007].

Online representation of places has been the object of several publications in the literature, looking at how we may navigate virtual representations as a structure for effective information retrieval. For example, [Zook and Graham 2007] considers the possibility of 'visiting' virtual representations of places and collecting from them the information we are seeking. Consistent with using concise representations for concepts and communication (e.g., Twitter) tags are also the possible novel and powerful basis for representing digital location, place, events and semantic descriptions [Rattenbury et al. 2007; Hollenstein and Purves 2012]. For example it is possible that these could support more advanced presentations of urban places [Cranshaw et al. 2012], such as conceiving a 'place' in terms of meanings, sense of attachment, and satisfaction provided by the people interacting with it [Stedman 2002; Tuan 2001]. Consistent with this authors [Colombo et al. 2013] have also presented a proof of concept for a keyword extraction methodology using online tips and reviews from a number of online sources. Digital representation of information and opportunities for activities available at places are made available notably through the use of online crowdsourced information, such as reviews [Goodchild 2007].

From the handset, mobile devices and location-based services now allow the monitoring of geographical positions in real time, thus moving from an initial adaptation of online maps and navigators towards services more oriented to provide reviews and personalised recommendations such as *Yelp* and *Qype*[2]. Other services combine location and user mobility information with a social networking component. Among those *Foursquare*, *Flickr*[3], and *Google+ Local*[4] have converged to a place representation that focuses on the individual needs of users, in terms of users being at a particular location at a particular time often making use of tags, annotations and other user generated content [Hollenstein and Purves 2012], thus differentiating from an initial represen-tation of places as a stand-alone virtual location that appeared distant from the real user's needs [Tuan 2001; Zook and Graham 2007].

[2] http://www.yelp.co.uk/, http://www.qype.co.uk/
[3] https://foursquare.com/; http://www.flickr.com/
[4] http://www.google.com/+/learnmore/local/

While very effective, these commercial systems are generally "closed", centrally organised and cannot be guaranteed to be adaptive to the wider population. Beyond these approaches, *collaborative tagging*, that is the use of shared keywords and 'tags' as a form of metadata in content organisation, has been widely recognised as a basis for modern suggestion and recommendation systems [Golder and Huberman 2006]. Free tags, as opposed to controlled vocabularies are generally preferred by users as more personal and less cognitively demanding [Quintarelli 2005; Sinha 2006]. Location services can be used to mine crowdsourced data in a place and reusing keywords extracted from these descriptions (tags) to inform the user, use of this technique of summarisation is not yet widespread.

While most of this research assumes fixed data sharing platforms such as Online Social Networks, this paper's focus concerns a fully decentralised approach where devices interact to create collective knowledge about places without centralised mediation. In Section 3 we describe the model for tag creation and its structure. In Section 3.2 we describe the way in which devices interact and how information is exchanged. In Section 4 we present dissemination results for tags.

## 3. COGNITIVE CONCEPTS FOR TAG ORGANISATION

There are many potential ways to acquire tags, such as from machine automated approaches, extraction from recommendation systems through to direct human filtering and creation. Whatever the means of their creation, our focus concerns their organisation for crowdsourcing. In this particular crowdsourcing scenario, we are adopting the *prosumer* model, where the individuals both provide tags to others and consume them. In addition, the places themselves can have fixed devices that operate in exactly the same way as mobile devices, and which effectively provide a collective memory for that location. We address these issues by tackling two main challenges. Firstly we consider how each device, or node, memorises tags so that they can be accessed and navigated giving priority for relevance to the individual. This needs to be achieved so that tags are referenced to locations and vice versa. Secondly, in the face of a potentially diverse and large number of tags in each individual device's memory structure, it is important to identify ways in which tag exchange can be prioritised between a given pair, upon encounters. In the following we separately address these two challenges. In the rest of this section, we make use of the symbols reported in Table I.

### 3.1. Memory

To address the first challenge, we take inspiration from the cognitive associative memory description of the human brain [Anderson and Bower 1973; Raaijmakers and Shiffrin 1981]. Associative memory for humans is dependent on structuring concepts by relations and is often referred to as the class of *Semantic Associative Network* (SAN) models [Gawronski and Payne 2010]. These models focus on the patterns and strength of associative linkages among concepts in the brain. Semantic associative network models represent memory as a graph, with persistent concepts being the nodes. Where an association is made between concepts this is defined by an edge. Edges of a SAN are weighted, where the weight reflects the strength of each association in memory and the ability to recall the adjacent concept given the stimulating trigger of the neighbour.

To define a SAN for our physical domain, we represent the memory or perception of a physical location by a user $u$ at time $t$ by a *weighted graph* $G_{u,t} = (V, E)$. Vertices represent the tags describing locations that are known by $u$ and edges denote a connection that $u$ has observed (recently) between two tags, with a weight giving the strength of this connection at time $t$. Moreover, we consider that each node (i.e., tag) of the graph $G$ is annotated with the set of physical places associated to that tag as known by the device, illustrated in Fig. 1. For purposes of creating a singular memory at a location, devices representing physical locations maintain only the associations with their own name. In contrast, along with the association with the names of physical locations, mobile devices keep track of the number of times this association has been observed in exchanges with other peers.

Table I. List of Symbols

| *Symbol* | *Definition* |
|---|---|
| $G_{u,t} = (V, E)$ | *Semantic Associative Network* (SAN) of a node $u$ at time $t$ |
| $m_u(e_{ij}, t)$ | *Memory weight function*: models the change of weights of edges in $G_{u,t}$ |
| $\beta_u$ | "Speed" of the forgetting process |
| $p_{ij}^{u,t}$ | Popularity of edge $e_{ij}$ in $G_{u,t}$ |
| $M_{min}$ | *Memory weight threshold*: below this weight, an edge is removed from $G_{u,t}$ |
| $G_{d,t} = (V_d, E_d)$ | *Donor Network*, i.e. SAN of a device that selects information to pass to another node |
| $G_{r,t} = (V_r, E_r)$ | *Recipient Network*, i.e. SAN of a node that receives information |
| $C = (V, E)$ | *Contributed Network*, i.e. information passed from a *donor* to a *recipient* node |
| $w(e_{ij}, n, t, t')$ | *Retrieval weight function*: used to select the information to be exchanged |
| $\tau$ | Regulates the dependency of the retrieval weight on the communic. duration |
| $W_{min}$ | Minimum retrieval weight |
| $\theta_{rec}$ | *Recognition threshold*: edges seen more than $\theta_{rec}$ times are relevant |
| $T_{max}$ | Max. number of tags a node can exchange during an encounter |
| $L_{max}$ | Max. number of annotations per tag allowed at each information exchange |

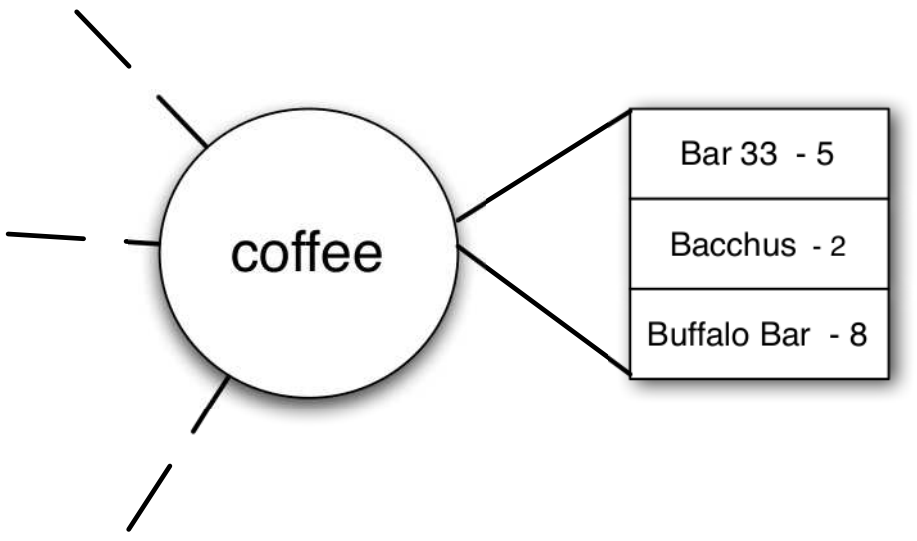

Fig. 1. Example of an annotated vertex in a device's semantic associative network (SAN)

*3.1.1. Recall Degradation.* Memory degradation is the process used by nodes to prune out from their memories the information that is believed to have become less relevant. This process allows memory space to be saved and, at the same time, lets nodes to be more adpative with respect to potential changes in the environment, allowing them to remove information that could not be useful in a changed context.

To model the degradation of memory between reinforcement, the edge weights in $G_{u,t}$ are defined by an exponential *memory weight function*, $m_u : (e_{ij}, t) \rightarrow [0, 1]$. The weight is set to 1 at each interaction that links the two concepts $i$ and $j$, and the rate at which the weight decreases is governed by the number of times the relationship has been observed. More formally:

$$m_u(e_{ij}, t) = e^{-\beta_{ij}(t-t')} \qquad (1)$$

where

$$\beta_{ij} = \frac{\beta_u}{p_{ij}^{u,t}}$$

$t'$ is the last time the relationship between $i$ and $j$ was observed before $t$, $\beta_u \geq 0$ is a factor regulating the speed of forgetting, and $p_{ij}^{u,t}$ is the number of times $u$ has observed the relationship up to time $t$, which we term it's *popularity*.

Note that at the initial observation of a link $e$, $t = t'$, and hence $m_u(e, t) = 1$. Furthermore, by setting $\beta_u = 0$ whenever $u$ represents the node at a location, and assuming links formed between all tags known by a node at time 0, the SAN a location stores about itself remains constant (i.e. unforgotten). Note that in the tag network stored at a physical location, edges still represent an association between tags, although clearly not bound to any human reasoning process, but representing how the tags representing that location are linked to each other. In Section 4 we explore the performance of the dissemination system starting from different ways of organising the individual locations' tags networks.

The exponential forgetting function is a well-known representation of the forgetting process in cognitive psychology [Wixted and Ebbesen 1991]. Rather than a limit, the forget process helps human brains to discard less relevant information when making decisions [Schooler et al. 2005]. For this reason, whenever the value of the *memory weight function* for an edge $e_{ij}$ goes below a *memory weight threshold* $M_{min}$, $e_{ij}$ is removed from $G$. Since human memories are more likely to drop information that is rarely accessed than frequently used data, we bind the definition of the *memory weight function* to the edge popularity. Therefore, connections between tags that are rarely used during exchanges with other devices or locations are more easily "forgotten". The forget process also affects the retention of tags in memory where a node in $G_{u,t}$ is dropped from the graph in the case where it becomes disconnected due to the deletion of one or more of its outgoing edges.

An example of the forgetting process is shown in Figure 2. In this example, the edges drawn with thicker lines represent weaker connections between tags, whereas bolder lines indicate stronger relantionships. In the example, the initial connections shown in Fig. 2(a) between vertices are not renewed for a while, making them weaker, as shown in Fig. 2(b). Since they remain unused for another time lag, the weight of two of the edges becomes so little that they have to be removed from the node SAN (Fig. 2(c)). As a consequence, the vertex with the label *"coffee"* is now disconnected from the rest of the SAN. Therefore, it is deleted, as displayed in Fig. 2(d).

### 3.2. Information Selection and Exchange

We consider that, upon meeting, a node (either mobile or physical) starts to exchange its knowledge with the other party by selecting the most relevant concepts for that given contact at that time from its SAN. The information exchange process starts from concepts that the two parties have in common. From those starting points, the information is selected by a process similar to a sequential search over human associative memories [Wyer Jr 2007]. This search mechanism starts with an activated semantic concept (*key-concept*) and then proceeds vertex by vertex in the SAN, following the links that connect them. Whenever a "dead end" is found, the search is reinitiated. In order to determine the relevant paths to follow when exploring a SAN, we apply the Fluency Heuristic [Schooler et al. 2005]: when a node in a SAN has more than one outgoing edge, the link with the "strongest activation" (measured by the retrieval weight index explained in the following) is selected.

In order to better understand this mechanism, we exemplify it by using a high-level description of the information selection and exchange processes. The precise details of the algorithms are given later in this section. In the following example, we refer to the scenario presented in Fig. 3. We describe how a device (Device 1, in the figure) selects the most relevant information to be exchanged from its SAN, upon meeting with another node. Like the example reported in Fig 2, edges with bolder lines have higher weights in the device's memory.

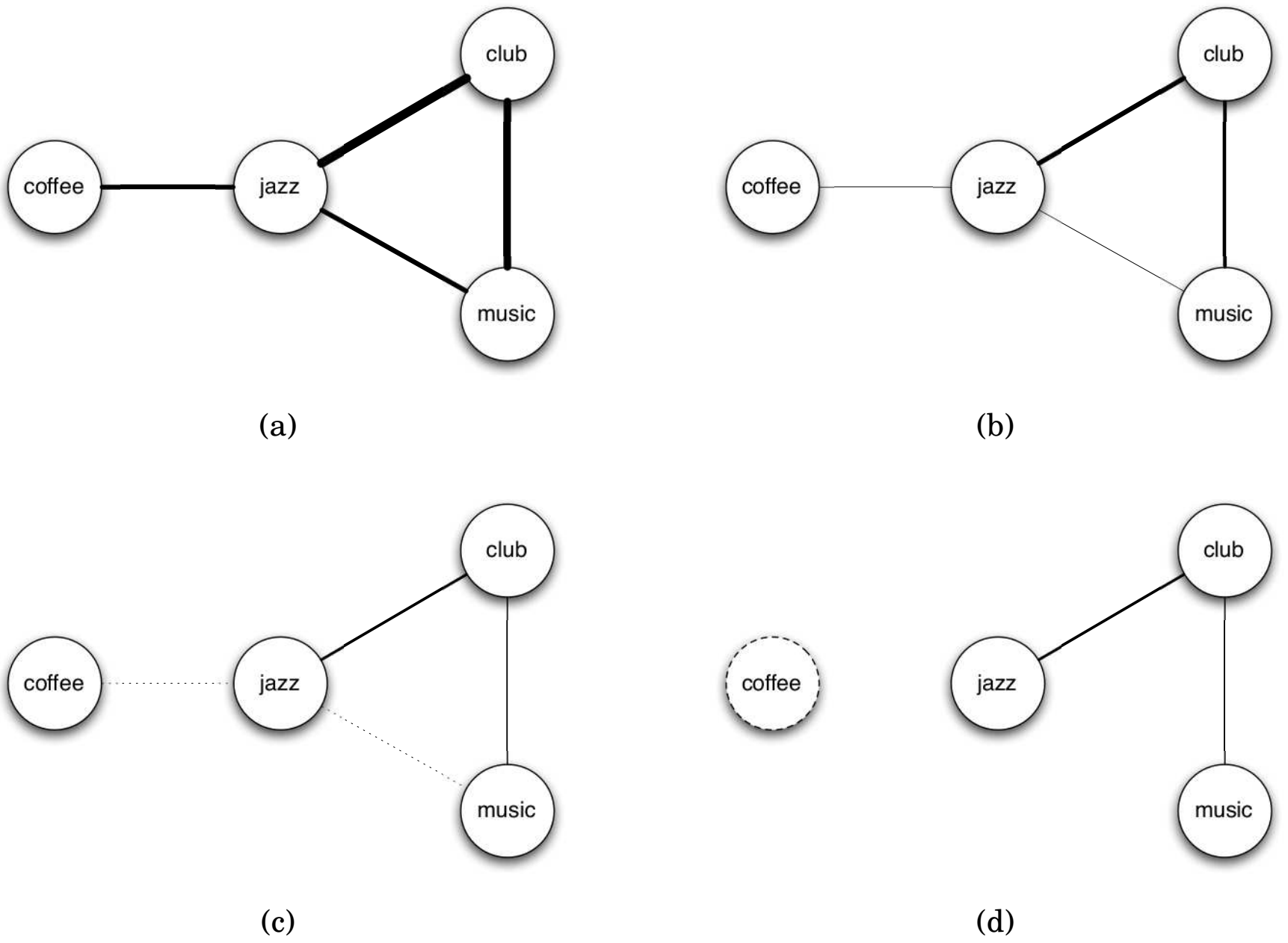

Fig. 2. An example of the *"forgetting"* process

In the example, the SANs of the two devices share a common tag, **A** (Fig.. 3(a)). Starting from this vertex, Device 1 explores its own SAN, as shown in Figures 3(b) – 3(d). The exploration of the SAN is used by Device 1 to select the information (edges and vertices) to pass to Device 2. At each step of the exploration process, an edge is selected on the basis of the ability of Device 1 to associate the edge endpoint to the exploration starting vertex (**A** in the example). This evaluation is a combination of the edge's weight and the distance of its endpoint from the starting vertex. As shown in Fig. 3(d) – 3(d), Device 1 evaluate that it is not possible to continue the exploration from vertex **D**, since its neighbouring vertices are considered too loosely connected with the starting vertex **A**. Therefore, Device 1 tries to continue the exploration, using another outgoing edge of vertex **B**.

When a vertex is included in the exploration, some of its annotations are also included in the data structure that is later used to pass information to the other interacting device. Fig. 4 shows an example of this selection, where only one annotation (the most popular) is allowed to be associated with the selected vertex.

In the example, we assume that constraints on resource consumption restrict the number of vertices to be exchanged between devices to a maximum of four. This limit is reached when vertex **E** is included in the exploration. Note that whenever an edge is included in the exploration process, it acquires the maximum weight, since it is "refreshed" in the Device memory. All the explored vertices and edges are copied in a data structure that is used to pass the information to Device 2 (Fig. 3(e)). This device, in turn, merge the received information to its own SAN, as shown in Fig. 3(f). Note that also Device 2 assigns to the received edges the maximum weight.

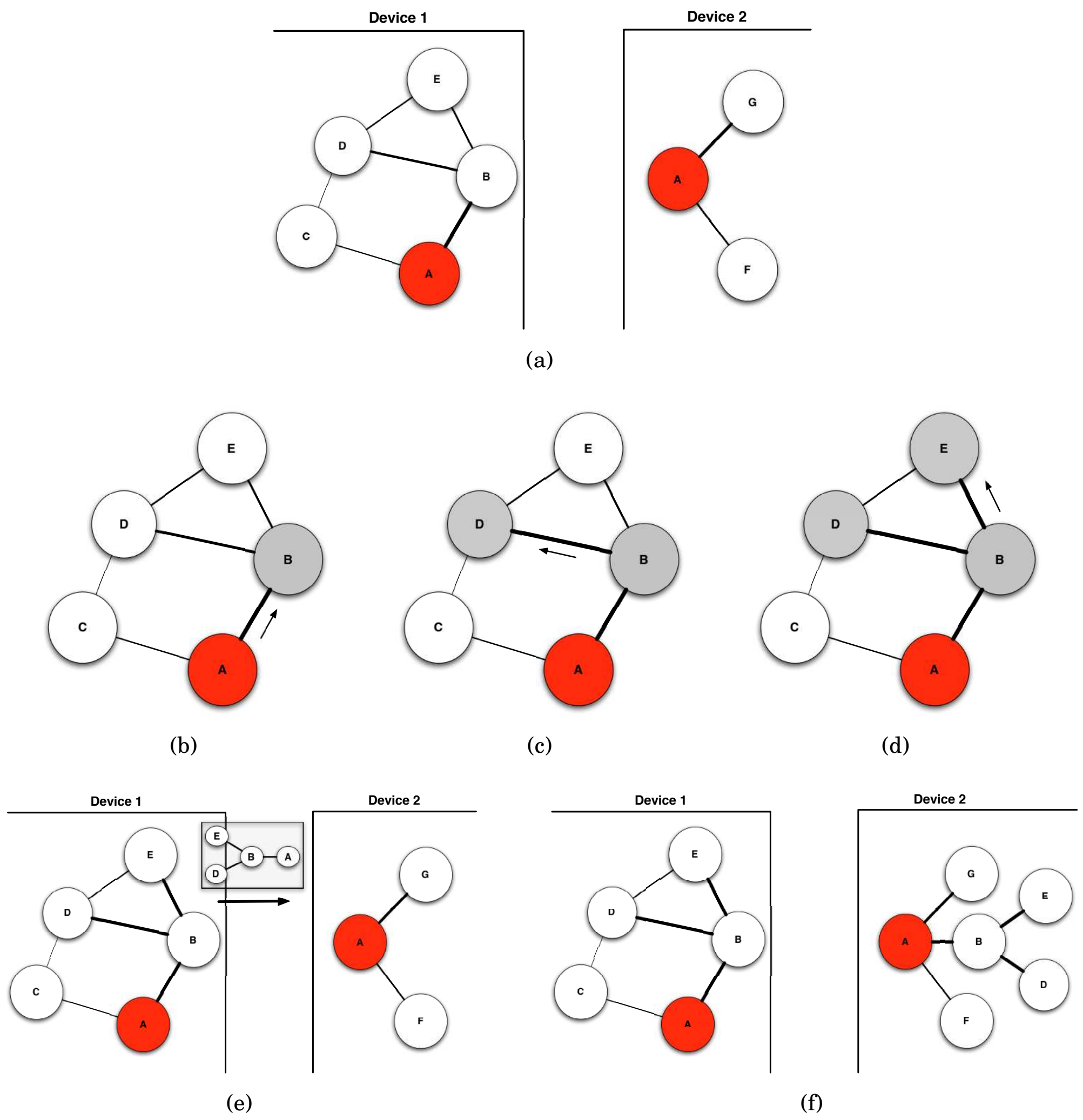

Fig. 3. An example of information selection and exchange

While Fig. 3 gives a high level description of the information exchange process, we now describe the precise algorithm by which the Fluency Heuristic is used by a physical location or a mobile node to retrieve the most relevant semantic information to be exchanged from its memory. In the following, the *donor* is the location or node that is selecting the information to pass to the *recipient* node. As stated previously, physical locations do not change their view of the tags which describe themselves. Thus, they can only act as donor nodes, and not recipients. On the other hand, two mobile nodes swap roles (donor and recipient) upon contact to realise a bidirectional exchange of information. We denote the SAN of the donor node as $G_{d,t} = (V_d, E_d)$ and that of the recipient as $G_{r,t} = (V_r, E_r)$.

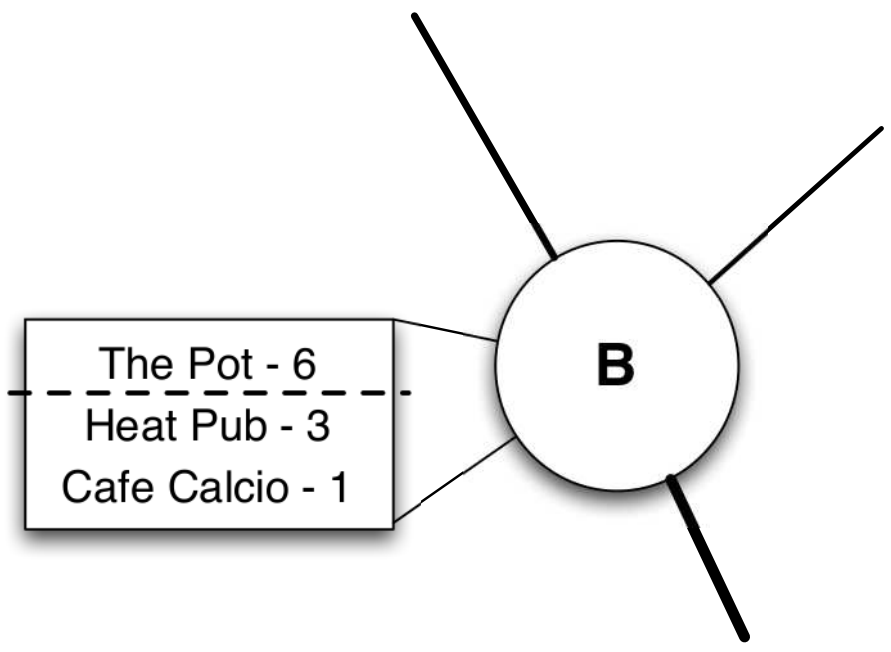

Fig. 4. An example of annotations selection

Based on the initial shared concepts between the two nodes (termed the *key vertices*), the donor selects a *contributed network*, $C = (V, E)$ as a subgraph of their SAN to be passed to the recipient based on their individual relevance. It is assumed that there exist resource consumption constraints limiting the number of exchangeable tags (i.e., nodes of the SAN) to a maximum value $T_{max}$. For the same reason, we assume that the donor node can select no more than $L_{max}$ names of physical locations associated with each tag it includes in the contributed network. The pseudo-code of the subsequent description is given in Algorithm 1 and 2.

**Algorithm 1** Compute Contributed Network from donor $d$ to recipient $r$ at time $t$

Let $G_{d,t} = (V_d, E_d)$ be the donor network;
Let $G_{r,t} = (V_r, E_r)$ be the recipient network;
Let $C = (V, E)$, $V = \emptyset, E = \emptyset$ be the contributed network;
Let $K = V_d \cap V_r$ be the set of *key vertices*
Let $S = \emptyset$ be the set of already visited *key vertices*
**for** each $v_{d,i} \in K$ **do**
    **for** each $v_{d,j} \in V_d - S$ **do**
        Increment popularity $p_{ij}^{d,t}$
    **end for**
    $S \cup = v_{d,i}$
    Let $rel(v_{d,i}) = \sum_{e_{ij}^d \in E_d} m_d(e_{ij}^d, t)$
**end for**
**for** each $v_i \in K$ taken in desc. order w.r.t. $rel(v_{d,i})$ **do**
    $C \cup = visit(v_{d,i}, 1, t)$
**end for**
Merge $C$ with recipient network: $V_r \cup V$; $E_r \cup E$

Taking the key vertices (sorted by relevance) one at a time, edges and vertices are visited and passed from the donor network to the recipient network using Algorithm 2, based on the Fluency Heuristic. Before exploring the SAN, since vertices are annotated with the physical location names, once a vertex is selected, its set of annotations is considered (line 5 of Algorithm 2). While physical locations always only associate their names to passed vertices, we assume that mobile nodes order vertex annotations in descending order with respect to their popularities. Thus, the locations that are most frequently associated with a given tag are chosen first to be exchanged. No more than $L_{max}$ locations are included in the annotations of the nodes included in the contributed network (lines 8–12). Then, the edges used to visit a SAN are selected using the Fluency Heuristic. Since the Fluency Heuristic favours *recognised* objects (i.e. objects seen more than a given amount of times) against unrecognised items, we start by excluding all the edges whose popularity is below a recognition threshold $\theta_{rec}$ (line 15 of Algorithm 2).

In order to replicate the subsequent discrimination made by the Fluency Heuristic and based on the perceived retrieval time, we consider that the most relevant edges are the ones with higher *memory weight* values and closer to a key vertex. The longer the contact time between two nodes, the more time is available to navigate the donor network and include edges and vertices in the contributed one. These factors are taken into account in the algorithm by computing a *retrieval weight* value for each outgoing edge $e^d_{ij}$ of a vertex $v_i$ (line 16). The *retrieval weight* is computed as in Equation 2:

$$w(e^d_{ij}, n, t, t') = m_d(e^d_{ij}, t)\frac{1 - e^{-\tau(t-t')}}{n} \quad (2)$$

where $t$ is the current time, $t'$ is the time at which $e^d_{ij}$ was last observed by $d$, $n$ is the number of hops in the shortest path to the nearest vertex in $K$ and $\tau$ is a "speed" factor that regulates the dependency of this value on the communication duration $(t - t')$. With a longer communication time, edges have more chances to be "warmed up" by the "discussion" and, then exchanged. We also refer to the *retrieval weight* as the *warm* value. Using this quantity, edges are sorted according to their retrieval weight and are then taken one at a time in descending order (line 16). Since $m_d(e^d_{ij}, t^*) = 1$ when $d$ represents a physical location, the retrieval weight for all incident edges in the SAN depends only on the distance from a key vertex and the communication duration. Each selected edge is included in the contributed network and allows the exploration of the donor network to continue (lines 4–23). Edges whose retrieval value is below a threshold $W_{min}$ are considered irrelevant for the information exchange and thus they are excluded from the contributed network (line 17). Moreover, once an edge is included in the contributed network, its memory weight is set to the maximum in both the donor and the contributed networks, since its inclusion in the exchanged data corresponds to an "activation" in memory.

Whenever $|V| = T_{max}$ and/or other edges cannot be selected from the donor network, the contributed network computation ends and the resulting graph is passed to the recipient node. This node, in turn, merges the received contributed network to the recipient one by simply adding any missing vertices and edges. Moreover, as for the donor node, all the edges received from the contributed network (new or already present) set their *memory weight* to 1, since they are "activated" by the "conversation". When merging the contributed and the recipient networks, the recipient node also considers the annotations of each vertex of the contributed network. In case the vertex was already present in the recipient network, the recipient node increases the popularity of all the annotations that already exist in the recipient network vertex and are present in the contributed one. When the vertex was not previously in the recipient network, or the annotation is not already present, the popularity of the received annotations is set to 1.

## 4. PROPERTIES OF THE SEMANTIC NETWORK OF LOCATIONS AT MOBILE NODES

In this section we present results about the properties of the information dissemination process for crowdsourcing the semantic associative networks (SANs) of mobile nodes within a simulated environment. We considered a $1km^2$ wide area where nodes move. In the results reported in Sections 4.3 and 4.4, we refer to the parametes shown

**Algorithm 2** Function $visit(v_i, n, t)$

Let $G_{d,t} = (V_d, E_d)$ be the donor network;<br>
Let $C = (V, E)$ be the contributed network;<br>
**if** $|V| < T_{max}$ **then**<br>
    Let $v_i = v_{d,i}$<br>
    Let $L_d$ be the set of the annotations of $v_{d,i}$<br>
    Order $L_d$ in desc. order w.r.t. the annotation popularity<br>
    Let $L = \emptyset$<br>
    **if** $|L_d| \leq L_{max}$ **then**<br>
        $L = L_d$<br>
    **else**<br>
        Construct $L$ from the first $L_{max}$ elements of $L_d$<br>
    **end if**<br>
    Annotate $v_i$ with $L$<br>
    $V \cup = v_i$<br>
    For each outgoign edge of $v_{d,i}$, $e^d_{ij}$, let $t'_{ij}$ be the latest activation of $e^d_{ij}$<br>
    Let $N(v_{d,i}) = \{v_{d,j} \in V_d | e^d_{ij} \in E_d\}$<br>
    Let $R = \{v_{d,j} \in N(v_{d,i}) | p^{d,t}_{ij} \geq \theta_{rec}\}$<br>
    **for each** $v_{d,j} \in R$ taken in desc. order w.r.t. $w(e^d_{ij}, n, t, t'_{ij})$ **do**<br>
        **if** $w(e^d_{ij}, n, t, t'_{ij}) \geq W_{min}$ **then**<br>
            $v_j = v_{d,j}$<br>
            $e_{ij} = e^d_{ij}$<br>
            $V \cup = v_j$<br>
            $E \cup = e_{ij}$<br>
            Record the latest activation of $e^d_{ij}$ as $t$<br>
            $C \cup = visit(v_{d,j}, n+1, t)$<br>
        **end if**<br>
    **end for**<br>
**end if**<br>
Return $C$

in Table II. In these cases, we have 100 nodes move according to a random waypoint model. Inside the simulation area, there are 10 static physical locations (placed uniformly at random in the simulation area) that spread their information. In Sec. 4.5 we divide the nodes into three different social communities, using 33 nodes and 3 physical locations per community. In this case, nodes move according to a model of real humans moving patterns. The simulation parameters are reported in Table IV. In all the scenarios, the description of physical locations is derived from a real-world dataset, described in more detail in Section 4.2. These simulation settings have been chosen as they are able to highlight the *general* behaviour and the macroscopic features of the proposed approach, allowing us to give insights into system performance in a realistic scenario. In particular, we use synthetic mobility traces, since they allow to deploy different and complex situations and scenarios. This fact gives us the possibility to study, in a controlled way, the sytem performance and its sensitivity to the various parameters, allowing to reach a greater understanding of the system behaviour. Moreover, it is worth noting that traces generated are genrated with the HCMM simulator [Boldrini and Passarella 2010], that is based on models of real human spatial and social movements. In fact, in reference [Boldrini and Passarella 2010], the simulator has shown to reproduce quite closely the behaviour of real users' mobility observed in popular traces. Therefore, we consider the use of HCCM synthetic traces as a good trade-off between the needs to control mobility for our sensitiveness analysis, and realism of the used mobility traces.

With respect to these settings, we can make some general considerations about the behaviour of the system in case more physical locations, more tags and/or more users are present in a wider area. In particular, with more physical locations and/or richer descriptions (i.e. more tags), it is easy to imagine that the knowledge dissemination process will proceed more slowly, since more data is available in the environment. However, the dissemination process is designed to let nodes exchange between themselves only the most relevant data they hold, considering the knowledge already acquired by the other interacting party (the process starts from common key concepts). At the same time, physical locations pass to interacting devices the part of their descriptions that is believed to be the most relevant. Also in this case, the decision takes into account the knowledge already present in the node that is communicating with the location. This fact, coupled with the forget mechanism, forces information that is less relevant in the environment to be diffused with more difficulties. Nevertheless, this information stays alive in the network, as it is constantly stored by places. Therefore, it is ready to be diffused, should it become of greater interests for users. The net effect is thus that the system scales automatically, by allowing only circulation of information that is relevant for users, and, at the same time, self-adapts to the dynamically changing interests of the mobile users. On the other hand, when a greater number of devices (i.e. users) are present in the environment, each node will take advantage of an increased number of "sensors (i.e. the other devices), that explore and collect information about the actual context, thus easing the information spreading process. Additional nodes will thus contribute additional capacity to the overall network, by contributing their shared memory for the global dissemination process. Moreover, when a node moves in a wider area, we can assume that every time it departs from a given zone, it will stay away from it (and its physical venues) for longer periods. This fact is due to longer travelling times and the time needed for possibile explorations of new areas. The information collected in previously visited regions i) could be less relevant in a changed context; ii) and it will not be used again for a sufficiently long time. One of the goals of the forget mechanism (Sec. 3.1.1) is exactly to let nodes be more adaptive to changed contexts, letting them drop information that is considered to have become less relevant in a new situation.

In order to model the SANs of the physical locations, in the following we consider three different configurations, based on the tag popularities derived from the *TF-IDF* frequencies of the locations tags in the dataset. The first configuration organises the nodes of the locations' SANs as a chain, with the most relevant tag at one end, connected to the second most popular tag, and so on, till the least popular tag at the other end. Semantic networks derived from online description could be also viewed as the aggregation of the associations made by a plurality of different users. Studies in the cognitive sciences (e.g. [Deyne and Storms 2008]) report that aggregate semantic associative networks show scale-free properties. Therefore, we also use two more clustered approaches, obtained using the algorithm reported in [Holme and Kim 2002]. Since this is a growing model of a graph with scale-free properties, for each location we run the algorithm by introducing the vertices in the graph growing process on the base of their *TF-IDF* order of relevance. The first configuration has a clustering coefficient of about 0.2 and the other one has a clustering coefficient of 0.5. Hereafter, we refer to all these three configurations as the Chain, CC=0.2 and CC=0.5 configurations, respectively.

At the start of the simulation each node SAN is initialised by choosing a group of tags from the set of all the available tags. Each tag has a 0.01 probability to be added to a node SAN. Initially, tags in a node SAN are not connected to each other, i.e. the node SAN do not have any edge. Moreover, tags are not annotated. Nodes seek to acquire knowledge about relationships between tags and association to physical location through the interaction between physical locations and other nodes. The performance metrics that we use are the *Hit Ratio* and the *Coverage*. The Hit Ratio is the average over all the nodes of the ratio between the number of tags held by each node and the overall number of tags available from the physical locations. This indicates the amount of information acquired by the nodes in the system. The Coverage measure is defined as the average of the per-node Coverage. This value, in turn, is computed as average (computed over all locations stored in the node SAN as tags annotations) ratio between the number of tags in the node SAN annotated with a location, and the number of tags that describe that location and are also stored at the node. Note that the node may not store all the locations associated to a stored tag, and thus coverage measures how complete is the information for the stored tags. In order to make the values of the *memory* and *retrieval* weights more intuitive for the reader, we use a number of conventions. The notation $forget = 50$s means that the $M_{min}$ weight is set in such a way that edges with popularity 1 are dropped from a SAN in case they are not seen before 50s from the last time they were used in an exchange. On the other hand, the notation of $warm = 25$s means that the $W_{min}$ threshold is computed taking into account, as a reference case, an interaction between nodes of 2s, that let to include (*warm up*) at least edges at distance 1 from a key node if they are not used (i.e. they were subject to the forget process) from no more than 25s. All the reported results are the mean of 10 different runs of the algorithm, obtained by using 10 different mobility traces for the nodes and 10 different placements in the area for locations.

In the following, we first present an analysis of the performance of the proposed solution. In particular, we compare the system against another reference solution and we report how the degree distribution of the individual nodes' SANs is related with the complete information available in the environment. These are the main findings about the behaviour of our system among the ones reported in [Mordacchini et al. 2013a]. We refer the reader to [Mordacchini et al. 2013a] for a complete analysis of those perfomance figures. In this paper, we extend our investigation about the system behaviour by testing it under more dynamic and complex scenarios. In particular, in Sec. 4.4, we study how the proposed solution behave when using dynamic locations' SANs, while in Sec. 4.5 we investigate the results of the system when nodes are grouped into different communities.

### 4.1. Overview of the Key Results

In the results presented in this section, we highlight the main features of the proposed solution. In particular, we can show how it is possible to conclude that the exploitation of structured infomation, i.e. the SAN, allows the cognitive-based system to outperform an epidemic-like competitor approach, in terms of both the dissemination of tags (and the relantionships between them), and the associations of tags with the physical locations they describe (Sec. 4.3). Moreover, it is worth noting that , by using the cognitive-based approach, a node is able to organise the information received from the environment in such a way that its final SAN closely approximate some of the main features of the global information available in the whole system, like the vertex degree distribution.

The system shows this behaviour with both static and dynamic physical location descriptions. When physical locations are allowed to change their descriptions (Sec. 4.4) upon interaction with mobile devices, the dissemination of tags is slowed down. This is due to the fact that physical locations limit the dissemination of the less relevant tags in the environment. In fact, one of the results is that physical locations change their perception about the relevance of the tags that describe them. On the other hand, this dynamic process does not affect the dissemination of the associations between tags and locations. Moreover, nodes individual SANs continue to reflect the main features of the global graph of the available information.

Table II. Main simulation parameters

| *Simulation Parameters* | |
|---|---|
| Simul. Area | $1000m \times 1000m$ |
| Numb. of Nodes | 100 |
| Numb. of Phys. Loc. | 10 |
| Node speed | unif. in $[1, 1.86]m/s$ |
| Transm. range | 20m |
| Simulation time | 75000s |
| $\beta_u$ | 0.1 |
| $\tau$ | 0.1 |
| $\theta_{rec}$ | 5 |

The results above are obtained in scenarios where all the nodes move inside the same physical space and could potentially come into contact with all the physical locations. Other results (Sec. 4.5) assess system performance when nodes and locations are divided in physically separated groups, or communities. In this case only a small number of devices (the travellers) can move between groups, allowing the dissemination in the system of the information about locations placed in different areas. The proposed approach allows nodes to rapidly acquire tags and associations about locations in the communities they move in. In addition, node are also able to become aware of the description of locations placed outside the groups they never visit. This is valid for both travelling and non-travelling nodes. Under this scenario, a sensitivity analysis of the cognitive-based solution shows that the higher the forget threshold (i.e. the information is allowed to stay longer in the devices memories) the lower its impact on the performance, since other factors become more relevant in this case. Additionally, the way locations organise their internal descriptions has a lower impact in the acquisition of the descriptions of locations places outside the communities visited by a node. In fact, since this information can be spread only through the mediation of other devices, the original organisation become less relevant.

### 4.2. Creating Tags for Experimentation

In practice tags could be acquired from many diverse sources, including user generation of the web. For experimental purposes we have produced a data set for venue representation using tags. In this work we have considered 10 venues in the city of Cardiff (UK) using a ‘global’ corpus consisting of the collection of all documents considered. This results in a total of 2210 tags, with an average of 221 tags per location.

The methodology we use was presented as a proof of concept in [Colombo et al. 2013] and shown as example in Figure 6. This consists of a keyword extraction process from aggregated text obtained from online reviews that represent the perception of users about a place (venue), rather than its objective description. The procedure returns a weighted list of tags where each weight represents the ‘importance’ of specific keywords. As an example the words frequency in the text is a simple version. To generate tags we take a number of real world candidate venues and produce a document that aggregates text from online reviews, user tips and comments, and other keywords that can be found online. Google+, Yelp, Qype (for reviews) and Foursquare (for tips) are used as data sources. Using the text document for each venue as an input the procedure that extracts and weight a list of keywords is then executed. The end result is that each venue is represented by a $n$-dimensional vector $v$ with each dimension $v[i]$ being mapped to a distinct individual tag. Filtering is necessary to remove punctuation and so forth. This is achieved by applying a Natural Language Toolkit library [6] to filter by different parts of speech (POS) such as adjectives, nouns, verb and adverbs. This is also used to tokenise, un-capitalise, strip of punctuation, remove unwanted words such as conjunctions, stop-words, repeated words, non-english words etc.

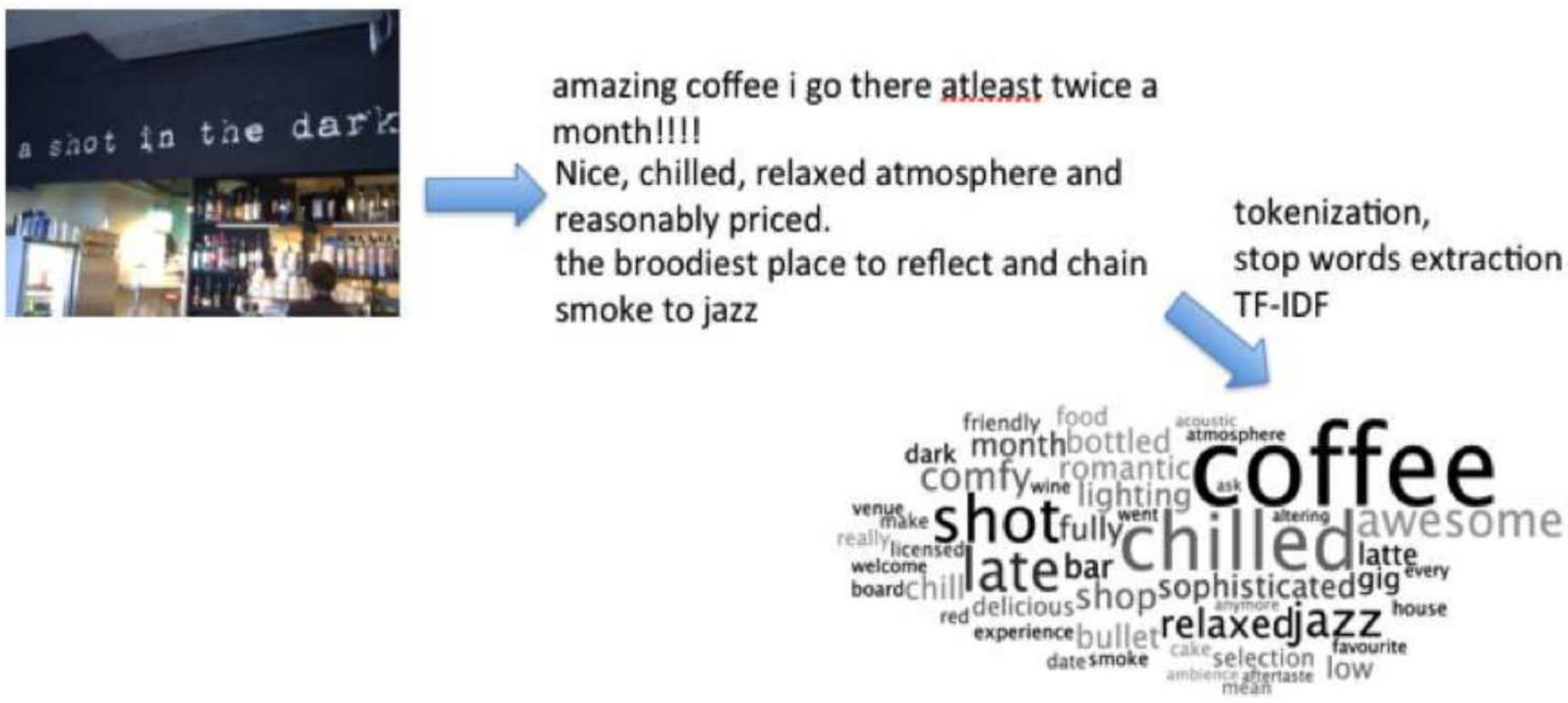

Fig. 5. A typical tag cloud

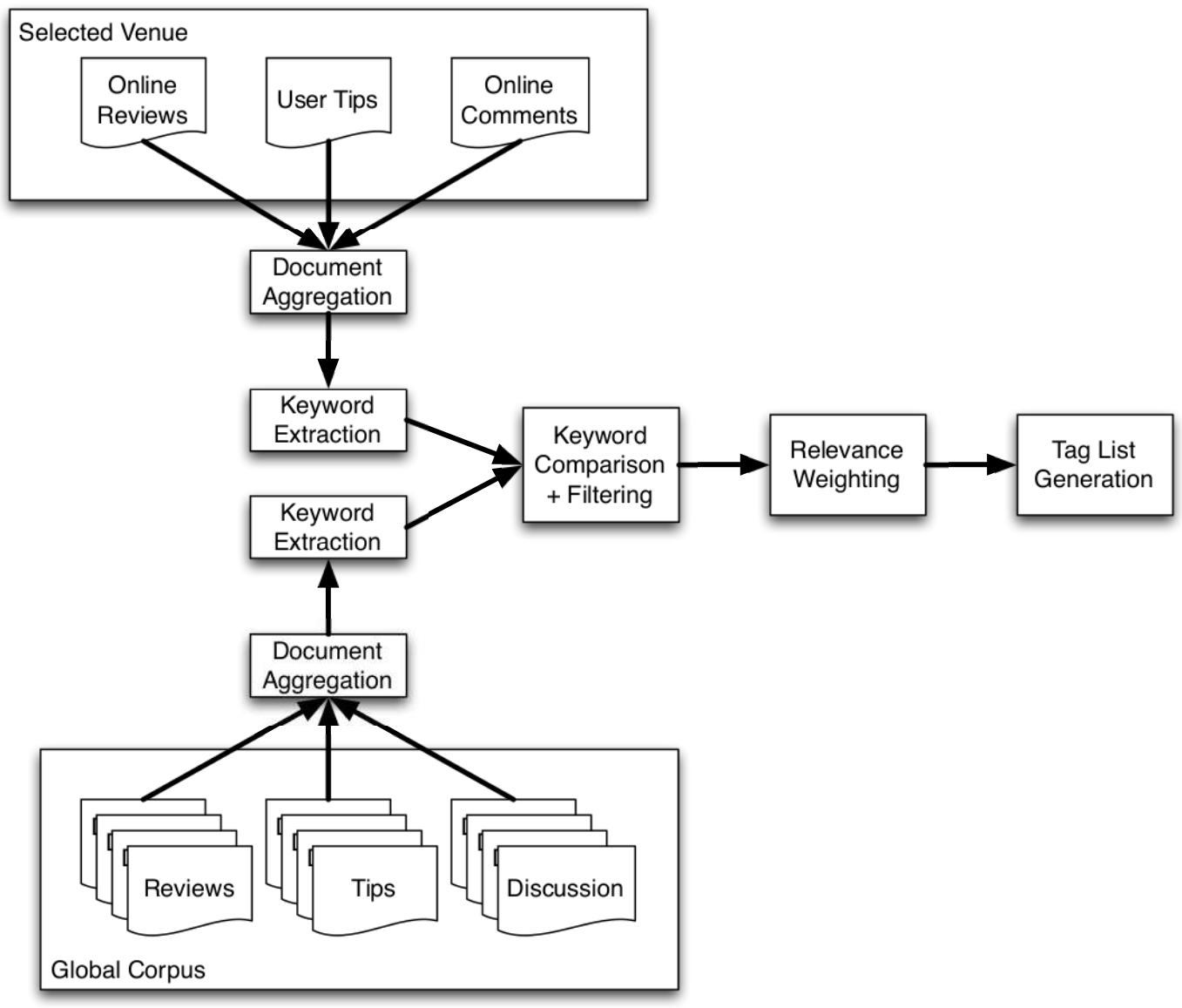

Fig. 6. Tag-list generation process

To produce the tag weighting we adopt TF-IDF (see [Manning et al. 2008]) that references the scaled popularity of the tag against a reference text of documents referred to as the *corpus*. Specifically the *TF-IDF* weighting scheme assigns to term $t$ a weight in document $d$ given by:

$$tf - idf(t, d) = tf(t, d) \times idf(t) \quad (3)$$

[6]http://nltk.org/

The *TF-IDF* weighting ensures that the highest weights are given when the term occurs many times within a small number of documents, and lower terms are given when the terms occurs fewer times in a document, or occurs in many documents. The lowest value occurs when the term occurs in virtually all documents

### 4.3. Comparison of cognitive vs non-cognitive solutions

In this first set of results, we show a comparison between our solution and an epidemic-like [Vahdat et al. 2000] data dissemination scheme. The simulation parameters are the ones shown in Table II. In the epidemic scheme used for comparison, the tags selected by locations and mobile devices to be passed to another encountered peer are chosen uniformly at random from the set of data they hold in their memories. For each of the exchanged tags, a set of known associated locations is also passed to the other party. Again, the choice of these locations is done by selecting them uniformly at random from the ones in memory. In this comparison, the epidemic scheme is subject to the same restrictions of the cognitive-based approach on data exchange. Therefore, no more than $T_{max}$ tags can be passed at each encounter and no more than $L_{max}$ locations can be associated with each exchanged tag. Moreover, we assume that data in the epidemic approach is subject to an aging process, similar to that of the cognitive case. Since the information stored by devices and locations in the epidemic scheme is not structured, i.e. nodes and locations store only tags and they are not aware of the relationships existing between them. As a result, the aging process acts directly on the stored tags. For each tag stored by a device in the epidemic scenario, we compute a "popularity" value in the same way as for edges in the cognitive approach. Then, we can define a *memory weight* function $m(d_i, t) = e^{-\beta_i(t-t')}$ applied to each data $d_i$. It is simply the cognitive *memory weight* function (Form. 1) applied to data items rather than edges. We use this function with exactly the same parameters as the cognitive function, i.e. the $\beta$ value and the same $M_{min}$ threshold, used to drop items from memory. This epidemic scheme is a very simple benchmark for our algorithm. In particular, it does not use any semantic representation of tags and any association between them. Epidemic is the most simple scheme that can be used to disseminate location information without using our cognitive based approach, but subject to the same resource constraints. Comparing our scheme with epidemic allows us to check that the former, under the same resource constraints, is able to achieve better information dissemination, thus resulting in a more efficient use of the available resources.

In the following simulations, we vary the maximum number of exchangeable tags $T_{max}$ and the memory weight threshold $M_{min}$ values, keeping fixed the minimum retrieval $W_{min}$ and maximum number of exchangeable vertex annotations $L_{max}$ parameters. In particular, $L_{max} = 2$ and $W_{min}$ is computed for a *warm time* $= 25$s.

The results of the comparison are reported in Figs. 7–8. In these figures, it is possibile to observe that the cognitive-based approach is able to outperform the epidemic solution in terms of both Hit Ratio and Coverage. This fact happens independently from the tag and forget parameters and the locations' SANs configuration used in the experiments. In particular, note that, in all cases the epidemic approach reaches a point where it enters a sort of oscillatory behaviour, where the values of the performance figures increase and decrease, floating around a stabilisation point. This behaviour is due to the effects of the forgetting process. In this region, the increase of the performance figures due to newly acquired tags is soon compensated by the drop of the oldest, least popular tags. Despite its initial rapid growth, the epidemic approach is not able to increase it performance above a stabilisation area, determined by the forgetting process.

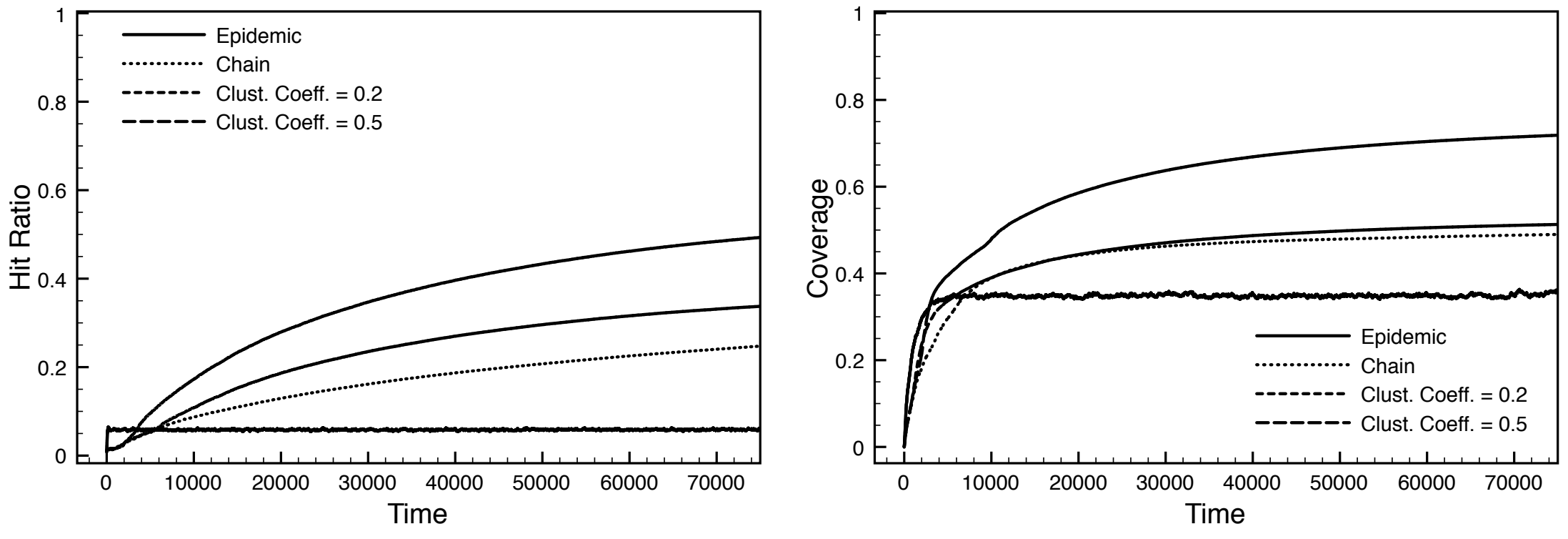

(a) Hit Ratio (b) Coverage

Fig. 7. Hit Ratio (a) and Coverage (b) results comparison; #tags =75, forget=75s

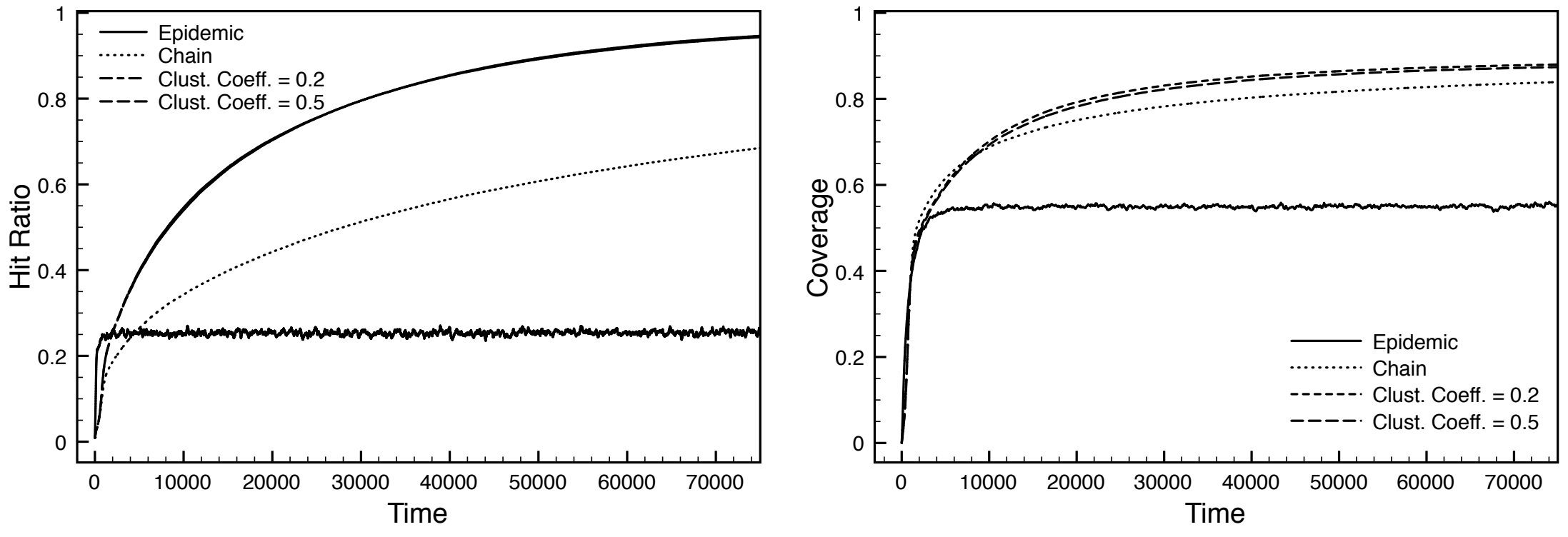

(a) Hit Ratio (b) Coverage

Fig. 8. Hit Ratio (a) and Coverage (b) results comparisons; #tags =150, forget=150s

On the other hand, the cognitive-based approach takes advantage of a structured information representation (i.e., SANs), where the forget process is applied to the edges in each SAN, before it impacts on the data (tags). Specifically, organising tags in semantic networks allow the system to avoid oscillations, as important tags have time to become "well established" in the nodes semantic networks, while tags that are really less important are correctly dropped.

Moving our attention to the performance of the cognitive-based approach, it is possible to observe that more clustered locations' SANs allow a better flow of information. This is an expected result. In fact, in more clustered SANs, tags are closer to each other in the locations SANs, and thus it is easier for each tag to be passed in the mobile nodes SANs. This is an effect of the process described in Algorithm 1. From its definition, it is possible to note that the lower the number of hops between a tag and the set of tags that are in common with another encoutered peer, the higher the chance of ending up in the contributed network. However, differences between the two clustered configurations are evident only in the experiment with the lowest $T_{max}$ and forget values (Fig. 7). It has to be noted that the Chain configuration achieves Coverage results very close to the clustered configurations when using higher forget and tag limit values (e.g. Fig. 8). Thus, even with the most difficult starting configuration, information spreads efficiently using the cognitive based schemes.

With Fig. 9 we compare the properties of the individual nodes SANs with those of the Locations SAN. The latter is the result of the union of all the SANs of the physical locations available in the simulation. Therefore, it represents all the information that can possibly be learned in the system and, thus, it is the asymptotic SAN toward which the nodes' SANs would tend in case of infinite resources.

Fig. 9 shows the average CCDF of the degree distribution of the nodes' SAN at the end of the simulation. The figure reports results for all the three configurations of the SAN of the locations, and for three different combinations of the forget and tag limits. In all the results, it is possible to see that the average nodes' degree distribution has the same slope as the Locations SAN. Moreover, with higher values of forget and tag limits, the nodes' curves are very close to the Locations SAN curve. This fact points out that the nodes are able to organise the information in their memories in such a way that it closely approximates the characteristics of the global information of the Locations SAN. In particular, the nodes' SAN have the presence of vertices with high degrees, even if with slightly lower probabilities than that of the global information. This is particularly relevant for the Chain configuration scenario, where devices acquire the information in the form of strings of tags. Even with the most difficult conditions for the information diffusion process, the nodes' SAN self-organise in such a way that we can find hubs that allow to both bridge the description of different physical locations and determine correlations between different concepts.

### 4.4. Dynamic Locations' SANs

In the experiments reported in the previous section, SANs associated to locations are static. More precisely, the weights of the links between vertices (i.e., tags) are never updated, and are set to a conventional initial value. In the following, we report results obtained by making the SANs of the locations more dynamic. Specifically, we use the forgetting function to make weights of the edges in each location's SAN decrease over time, in case they are not used, i.e. exchanged with users upon contact. With respect to users' SAN, in this case edges do not go down the $f_{min}$ threshold, after which links are broken. Physically, this would mean that the importance of a certain tag for a location can fade out, but that tag will never disappear, since tags represent features of a physical place. On the other hand, when nodes visit a location, the common nodes between the location and the user will be refreshed, and their links (in the location's SAN) will accrue weight, increasing their relevance. In this scenario, locations first derive their descriptions (i.e. tags) from the cyber-world (i.e. comments on on-line sites, as reported in the dataset). Then, they adapt their perception of the relevance of their own tags as a result of interactions with the devices of the users that physically visit them.

Clearly, the first impact of this dynamic and adaptive behaviour is on the nodes' Hit Ratio and Coverage metrics. In fact, tags selection on the locations SANs is guided by their relevance, and it changes over time due to the interactions with mobile devices. Figures 10(a) and 10(b) present the effects on the Hit Ratio and Coverage metrics, respectively. We can observe that the Hit Ratio (Figure 10(a)) is lower than that obtained, with similar parameters, in the case of fixed locations' SANs as examined in Figure 8. This effect is due to the fact that less relevant tags lose weight in the locations' SANs and, therefore, are more difficult to be exchanged.

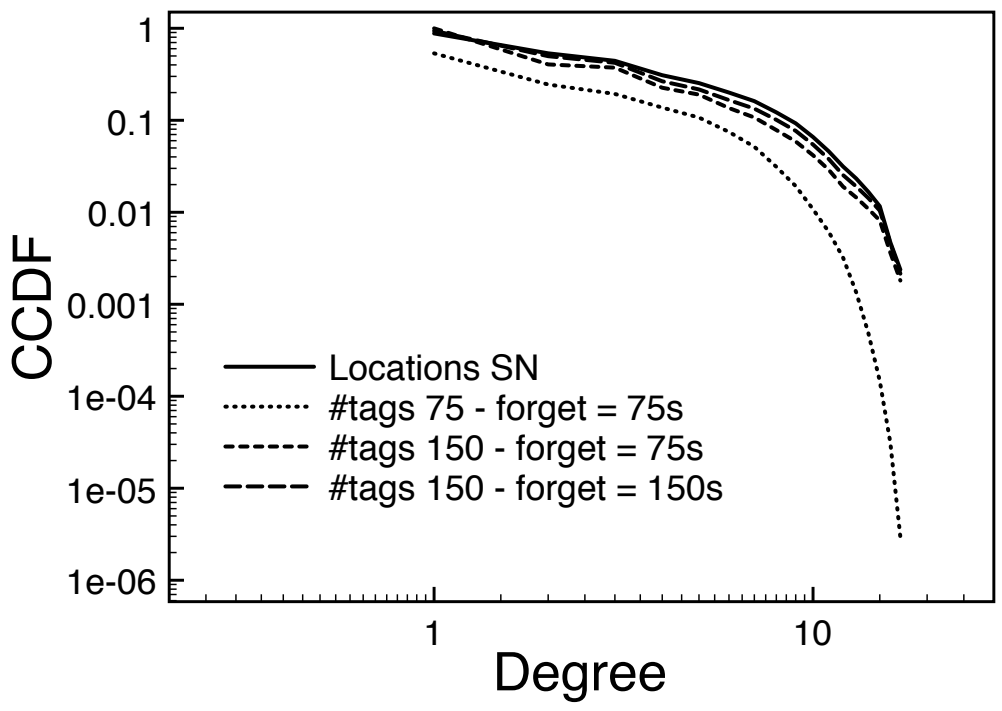

(a) Chain

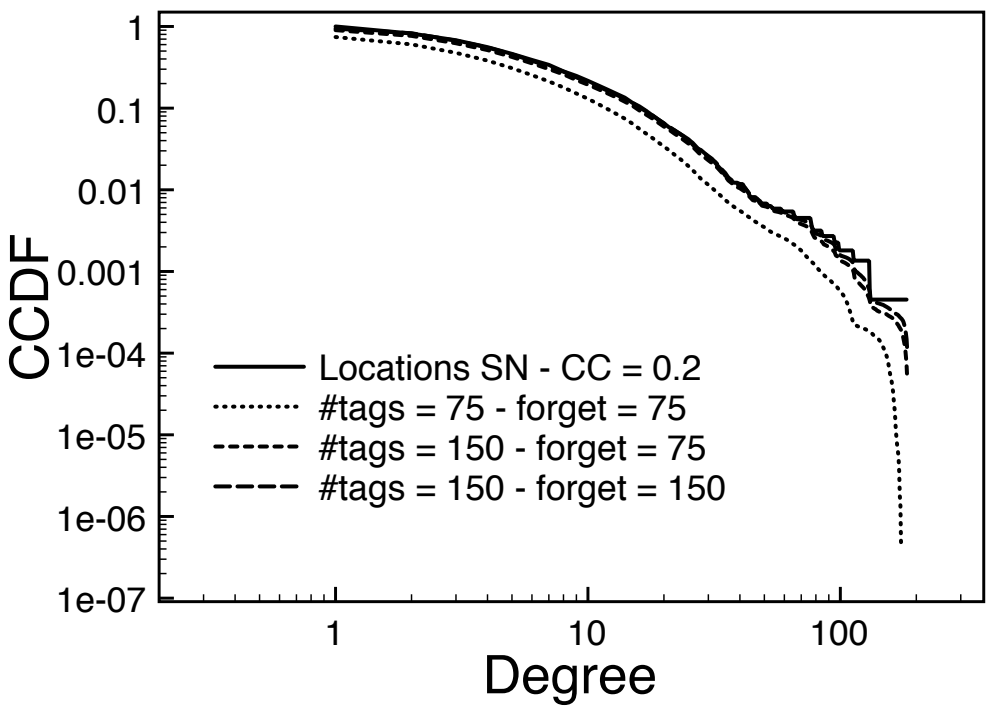

(b) Clust. Coeff. = 0.2

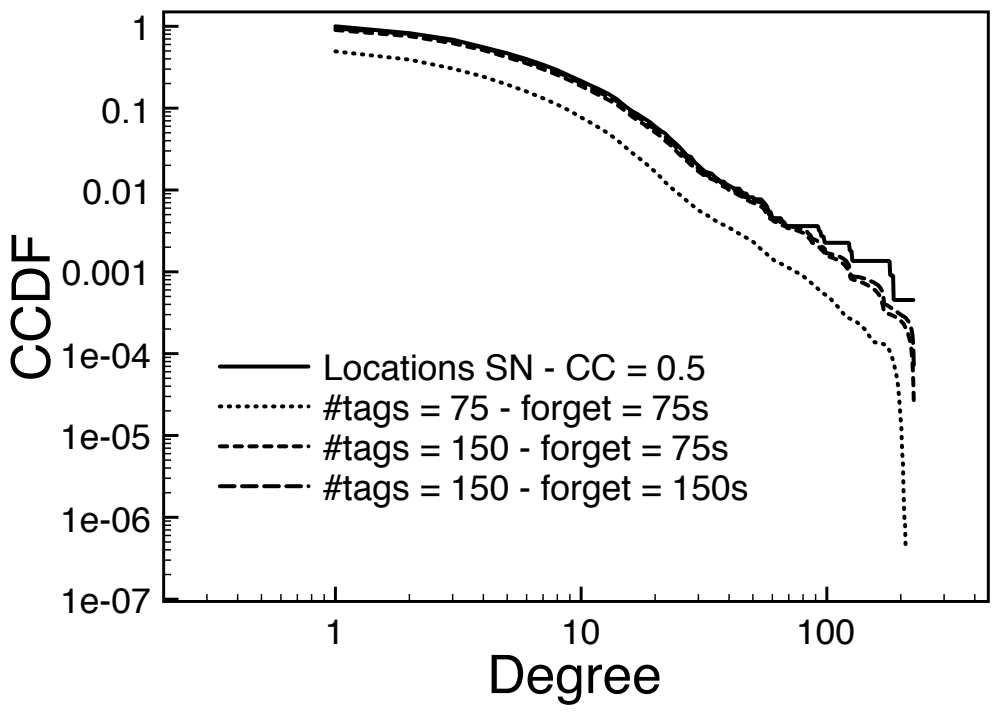

(c) Clust. Coeff. = 0.5

Fig. 9. Mean degree distribution in the nodes' final SAN

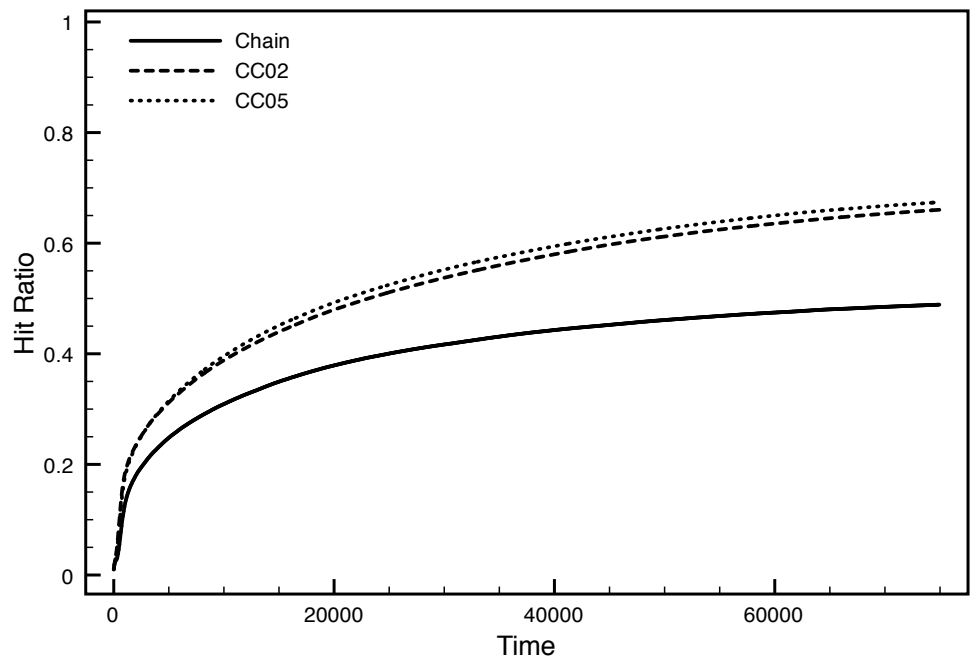

(a) Hit Ratio

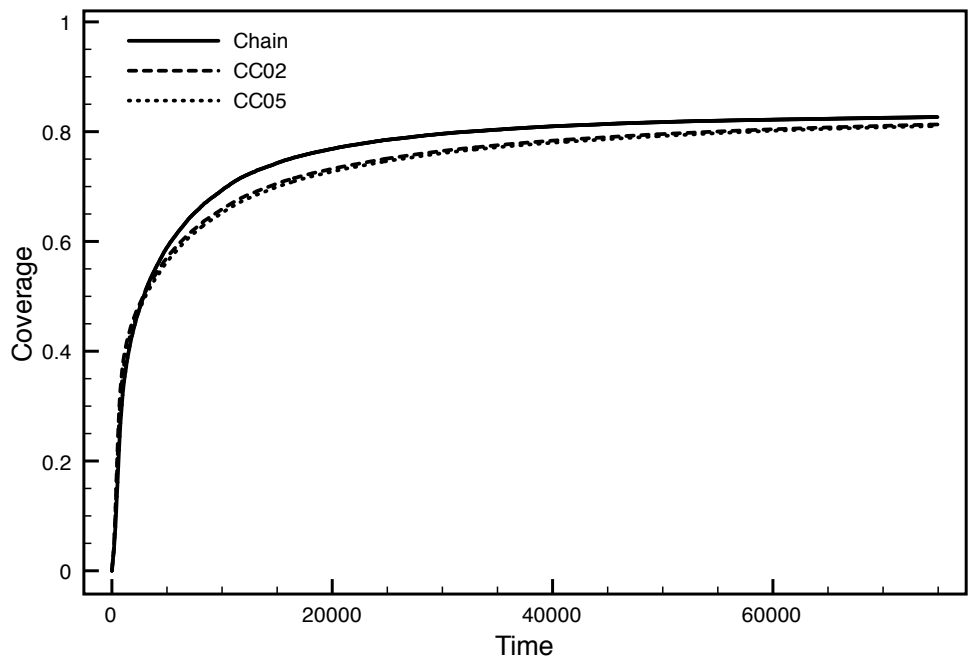

(b) Coverage

Fig. 10. Weighted degree Hit Ratio (a) and Coverage (b) results comparisons; #tags =150, forget=150s

On the other hand, the Coverage (Figure 10(b)) value is similar to that obtained with fixed locations' SANs as presented in Figure 8. With respect to this performance metric, the system achieves almost the same result under all the tested locations' SAN configurations. Therefore, while changing the locations' edge weights slows down the knowledge (i.e. tags) acquisition process, it does not impact the ability of nodes to associate the name of locations to the set of tags they carry in their SANs.

As for the static case, we investigate the properties of the nodes SANs at the end of the simulation. In particular, Figures 11(a), 11(b) and 11(c) (log-log scale in all the figures) compare the CCDF of the weighted degree distribution of the global locations' SAN and that of the SANs carried by mobile nodes. The global locations' SAN is computed by merging the SANs of all the locations. Whenever an edge appears in more than one SAN, its global weight is the average of the weights in each single SAN. The mobile nodes weighted degree distribution is computed as the average weighted degree distribution of all the nodes in the system. The behaviour of the system using locations' dynamic SANs is similar to that of static SANs. In fact, the slope of the nodes' CCDF curve is extremely similar to the locations' one, as for the static case (Figure 9). This fact confirms the ability of nodes to self-organise the information in their SANs in a way that reproduces the general properties of the global information available in the network.

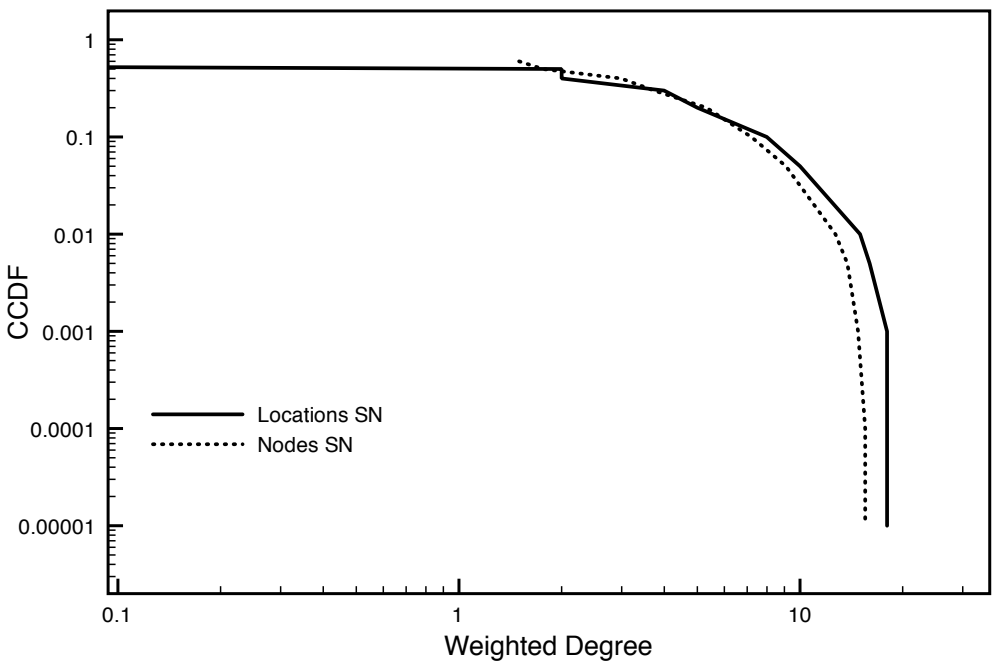

(a) Chain

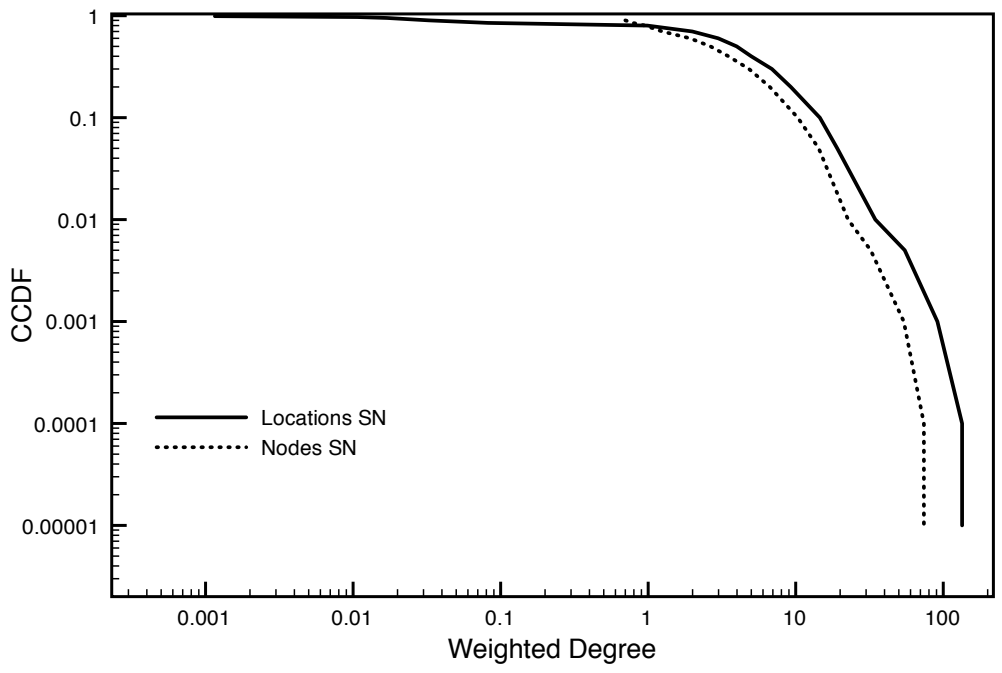

(b) Clust. Coeff. = 0.2

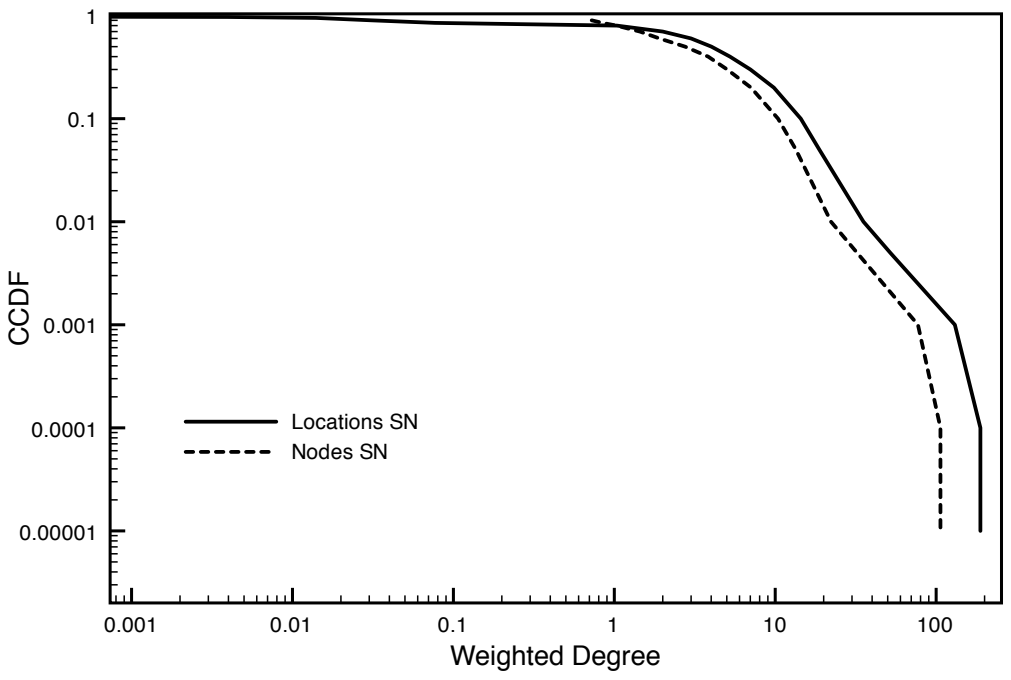

(c) Clust. Coeff. = 0.5

Fig. 11. Mean degree distribution in the nodes' final SNs

Table III. Kendall Tau-b Correlation Index

| Net. Type | $\tau_B$ |
|---|---|
| Chain | $0.8011 \pm 0.0084$ |
| CC02 | $0.6118 \pm 0.0492$ |
| CC05 | $0.6388 \pm 0.0531$ |

The previous figures presented the effect of dynamic locations SANs on the information diffusion process towards mobile nodes. We want now to investigate the effect of this dynamic configuration from the location's perspective. These results are summarised in Table III. Specifically, we measure the difference in the weighted degree rank of the vertices of the global locations SAN at the begin and at the end of the simulation. When the simulation starts, all the edges have the same weight, while, at the end, weights have changed as a result of the interactions with mobile nodes. Ranking vertices on the basis of the weight of their incoming edges, we can measure the difference between the initial and final ranks using the standard Kendall Tau-b correlation index. It is possible to observe that all the locations' SAN present differences between the initial and final ranks. Specifically, using the Chain configuration, the final rank is very similar to the initial one, since this configuration as fewer degree of freedom than the others. The test problem involving two locations' SAN clustering have final ranks that have more difference with respect the initial ones. These results highlight the fact that, using dynamic SANs, the fixed locations change the way they perceive the relevance of their information on the basis of the interactions that occurred with the environment.

### 4.5. Many Communities

In all the results shown in the previous sections, locations are supposed to be placed in a geographic area where each mobile device can move freely, possibly interacting with any other device or physical location. In this section we describe the results obtained in a scenario where nodes are grouped in different social communities. In this context, the simulation area is divided in cells of the same size and communities are initially assigned to different cells (referred to as *home cells*) and any physical contact among groups is avoided. Thus, the only way to exchange and obtain data among different communities is through node mobility. Each node can move in the cell of its community only (i.e., the home cells). The only exception are a subset of nodes in each community, named *travellers*. Each traveller is allowed to visit just one of the other groups. Travellers are the bridge that connect different communities, allowing the flow of information between them. In this scenario, each community has a traveller toward each of the other communities. Real user movement patterns in this scenario are simulated according to the HCMM model [Boldrini and Passarella 2010]. The HCMM model is a mobility model that integrates temporal, social and spatial notions in order to obtain an accurate representation of real user movements. The overall simulation parameter used in this scenario are reported in Table IV.

The results shown in Figure 12 are obtained under the Chain (i.e., the most difficult) configuration and with $forget = 35, T_{max} = 50, L_{max} = 2$ and $weight = 25$. Figures 12(a) and 12(b) show the average Hit Ratio and Coverage variations for non-travelling nodes. For each of the two performance figures, we present two curves, one showing the behaviour of the system with respect to the information locally available inside the node's community and the other for the data placed outside it. It is possible to observe that nodes inside the community are able to become aware of the space where they are moving (high Hit Ratio and high Coverage), while the information spread by travellers allows them the acquire knowledge about the description of the environment outside their home cell. Clearly, the knowledge acquisition process about data located in other communities proceeds more slowly, since travellers are the only source of information about the outside world.

Table IV. Simulation Parameters

| *Param.* | *Value* |
|---|---|
| Numb. of Communities | 3 |
| Nodes per Comm. | 33 |
| Locations per Comm. | 3 |
| Numb. of travellers | 6 (2 per comm.) |
| Total numb. of tags | 2098 |
| Simul. Time | 75000s |
| $T_{max}$ | 50 |
| $L_{max}$ | 2 |
| $weight$ | 25 |

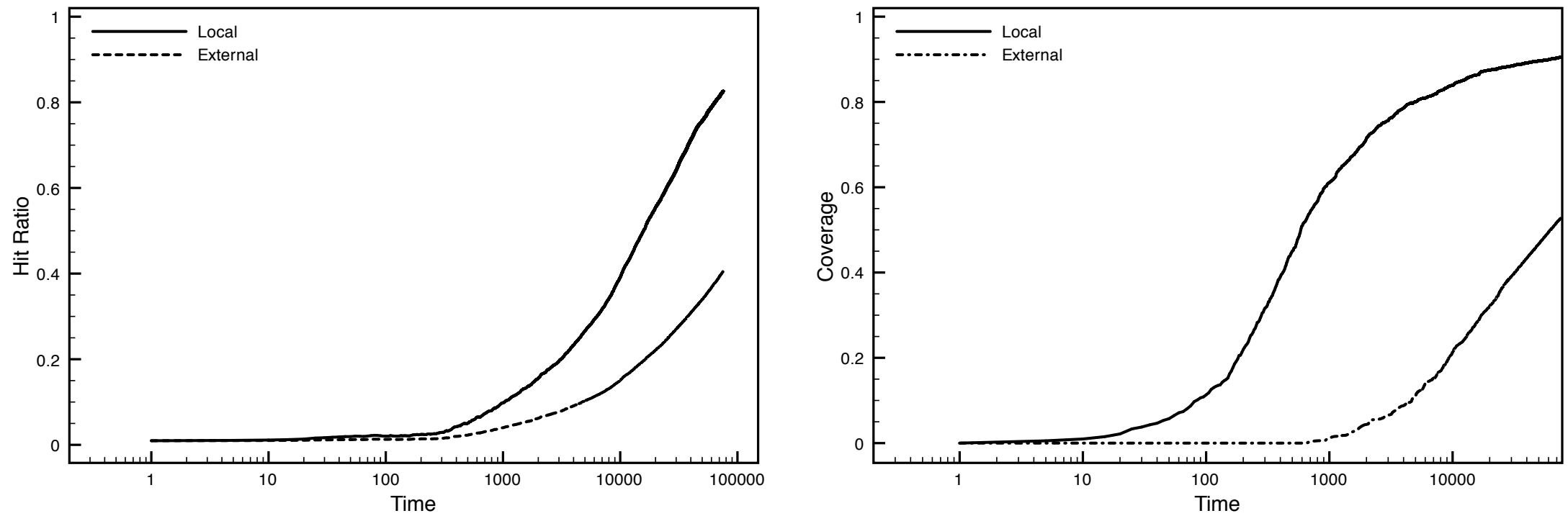

(a) Hit Ratio (b) Coverage

Fig. 12. Hit Ratio and Coverage of non-travelling nodes inside a tagged community w.r.t their own community and the external ones

In addition to the analysis of the behaviour of non-travelling nodes, we now give a closer look to the performance of travellers. Figures 13(a) and 13(b) present the Hit Ratio and Coverage performance of a tagged traveller. In this case, we have three separated curves: one for the traveller's home community, one for its destination community and one for the community that is not directly visited by the traveller. In this scenario travellers spend more time inside their home community rather than in the destination one. Thus, Hit Ratio and Coverage are higher for the home community than those of the other two communities. In addition, the Hit Ratio of the external community (i.e. the community not visited by the traveller) is slightly better than that of the traveller's destination group. This effect is due to the fact that information about the external community is spread inside both the groups visited by the traveller by other travellers to/from the external group. In contrast to the Hit Ratio results, the Coverage metric highlights that the association between the name of locations and discovered tags occurs more easily for the traveller, with respect to its home and destination communities. For locations that remain in the external community, this task is more difficult for the traveller, since it never comes in direct contact with that physical locations and, thus, it has to rely only on the information carried by other travellers.

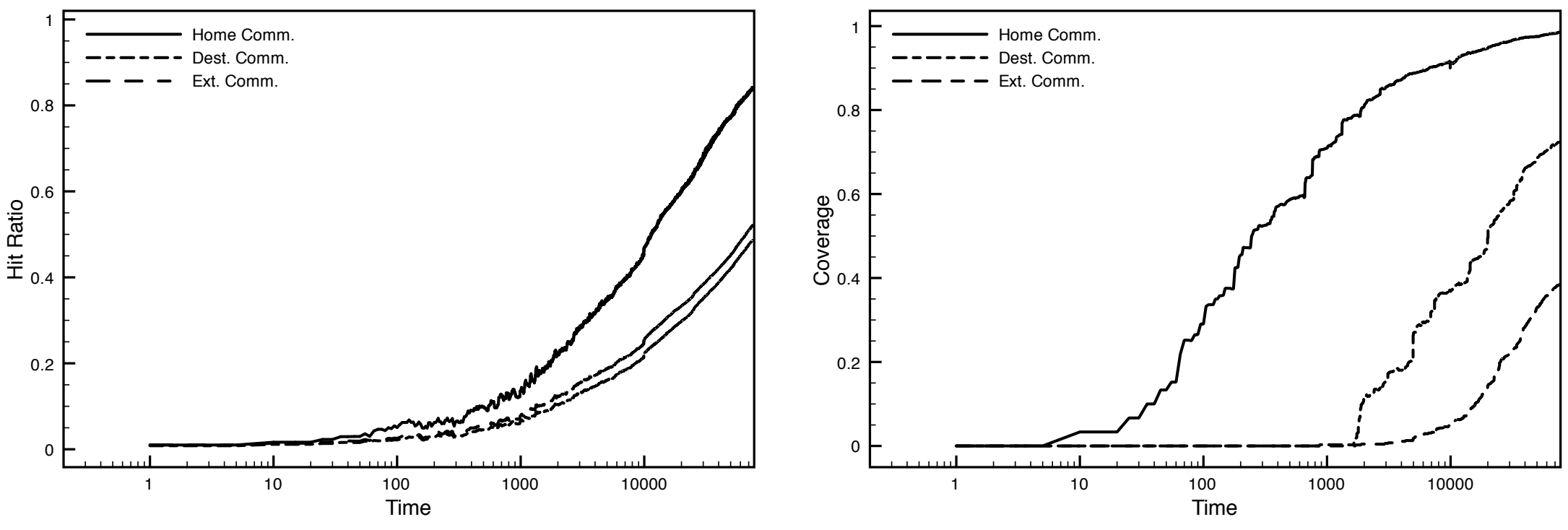

(a) Hit Ratio (b) Coverage

Fig. 13. Hit Ratio (a) and Coverage (b) of a tagged traveller w.r.t. its own community and the external ones

The previous results give us a first insight on the behaviour of the system in a multi-community scenario. In the following set of results, we investigate how the performance metrics are affected by the forget parameter and the locations SAN configurations.

Figures 14 – 16 present the behaviour of the system using three different *forget* values. As for the previous results, we present the average behaviour of nodes inside a tagged community and the performance of a tagged traveller. Figures 14(a) and 14(b) report the Hit Ratio of nodes in a tagged community related to the group's local and external information, respectively. Figures 15(a), 15(b) and 15(c) show the Hit Ratio of a tagged traveller for the data in its home, destination and external communities, respectively, while, for the same traveller, Figures 16(a), 16(b) and 16(c) report the results related to the Coverage metric.

In all these figures, it is possible to observe that the difference between the results obtained with $forget = 35$ and the ones achieved with $forget = 50$ is much lower than that existing between $forget = 35$ and $forget = 25$. This phenomenon can be ascribed to the fact that the higher the *forget* value, the lower the "marginal utility" of further incrementing it. This is because of the other settings of the system parameters. They become more relevant when the forgetting process attenuate its impact (i.e. for higher *forget* values) on the information dissemination mechanism. Therefore, with the other parameters unchanged, the relevance of the forget threshold diminish as far as it becomes higher.

Figures 17–18 report the behaviour of the system for a tagged traveller under the three different locations SAN problem, with $forget = 35$ and the other parameters set as per Table IV. Precisely, Figures 17(a)–17(c) show the Hit Ratio with respect to the locations in the home, destination and external communities, respectively. Figures 18(a)–18(c) show the Coverage of the traveller for the same communities.

Looking at the Hit Ratio performance, we can note that there is very little difference between the two clustered configurations (CC02 and CC05). Moreover, when using these configurations, the system performs better than when the Chain configuration

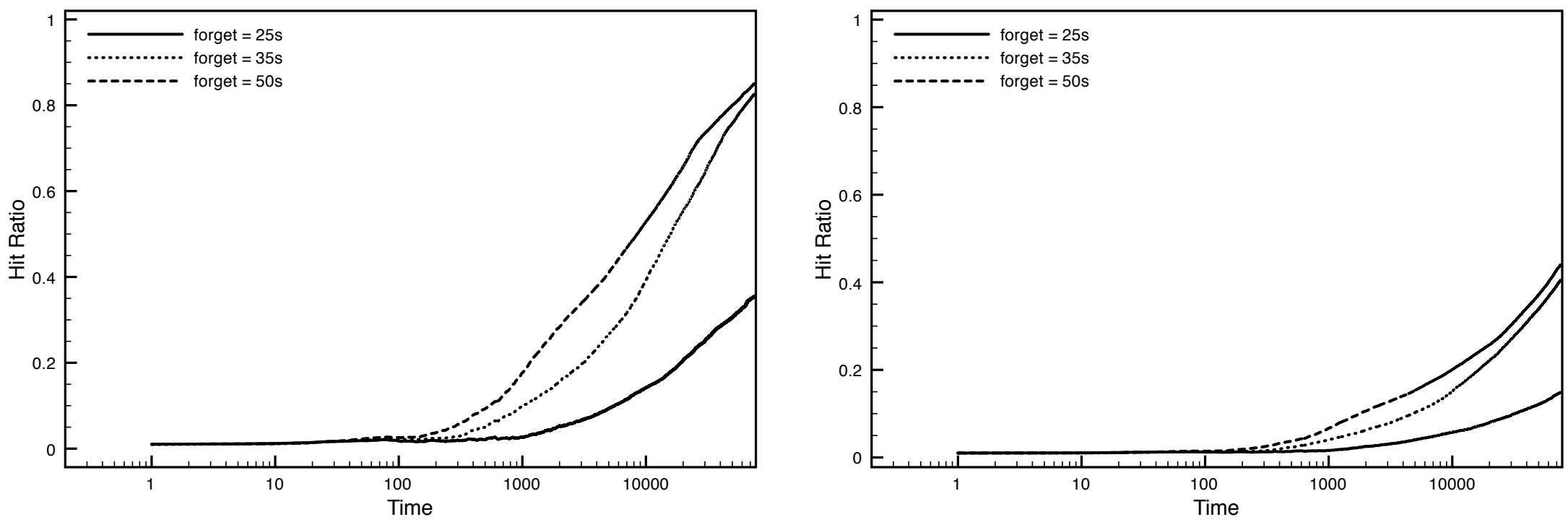

(a) Home Comm.

(b) Ext. Comm.

Fig. 14. Hit Ratio of non-travelling nodes inside a tagged community w.r.t their own community (a) and the external ones (b) for different values of the *forget* threshold

is exploited. This is consistent with the behaviour observed in a single-community scenario. However, the difference between the Chain and the clustered configurations is lower for the Hit Ratio related to the external community. This happens because the traveller becomes aware of that semantic information only through contacts with other mobile nodes. In fact, it does not acquire that data directly from the locations since they are placed outside the area where the traveller moves. Therefore, since the knowledge acquisition process is mediated by other nodes, the original SAN organisation of the locations has a lower impact.

Additionally the Coverage metric shows lower differences between the three locations' SAN organisation for the communities visited by the traveller, while there is a more relevant difference between the Chain configuration and the two clustered configuration. In this case, the traveller is able to learn the association between the semantic description and the locations by direct contact with the locations themselves. When making the *contributed network*, locations associate their label to each vertex. Thus, it is easier for a traveller to add the name of a location to the tags stored in its memory. In the case of the external community, it has to rely on contacts with other peers. These nodes are subject to limits in the number of locations that can be associated to a tag in the contributed network (i.e., the $L_{max}$ parameter). Only the most popular locations associated to a tag can be passed along with the *contributed network*. Hence, this process could be influenced from the configuration of the SAN in the locations from where other travellers first learned the association, before spreading it in other communities.

## 5. CONCLUSION

This paper has addressed the way in which crowdsourcing about places can occur through cognitive opportunistic networks. Cognitive overlay schemes for semantic memory have been introduced on top of opportunistic networks, resulting in a new mechanism to share and relate information about particular places. This is highly relevant to crowdsourcing knowledge about places. In this paper we have particularly looked at the underlying dynamics of dissemination within the cognitive overlay network, and the effects of the local memory structure on the dissemination. Under limited resources, it results in more effective dissemination of information with respect to reference solutions that do not exploit cognitive models. In addition, we have shown that a cognitive-based solution is very effective in disseminating data (under limited resources) also in dynamic conditions where information about places change over time, and users move according to different social groupings. Overall our results show that that using the same scheme that drives behavioural efficiency in terms of memory for the human brain is an effective structure that can be explored in the resource constrained opportunistic environment where crowdsourcing can be harnessed. This analysis of processes that support crowdsourcing paves the way for further study on the semantic knowledge base that could support future smartphone applications for crowdsourcing.

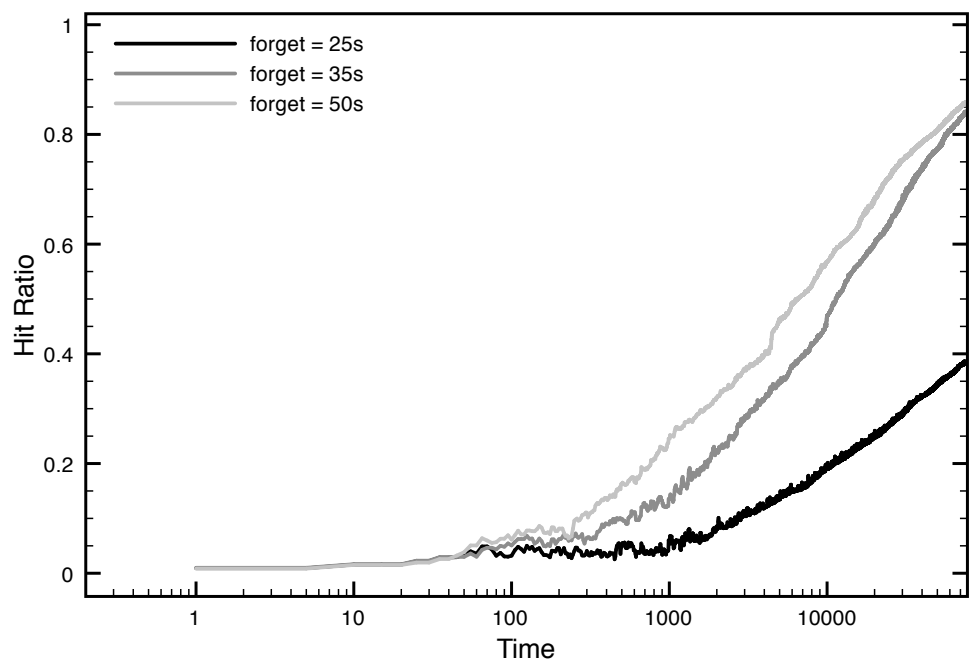

(a) Home Comm.

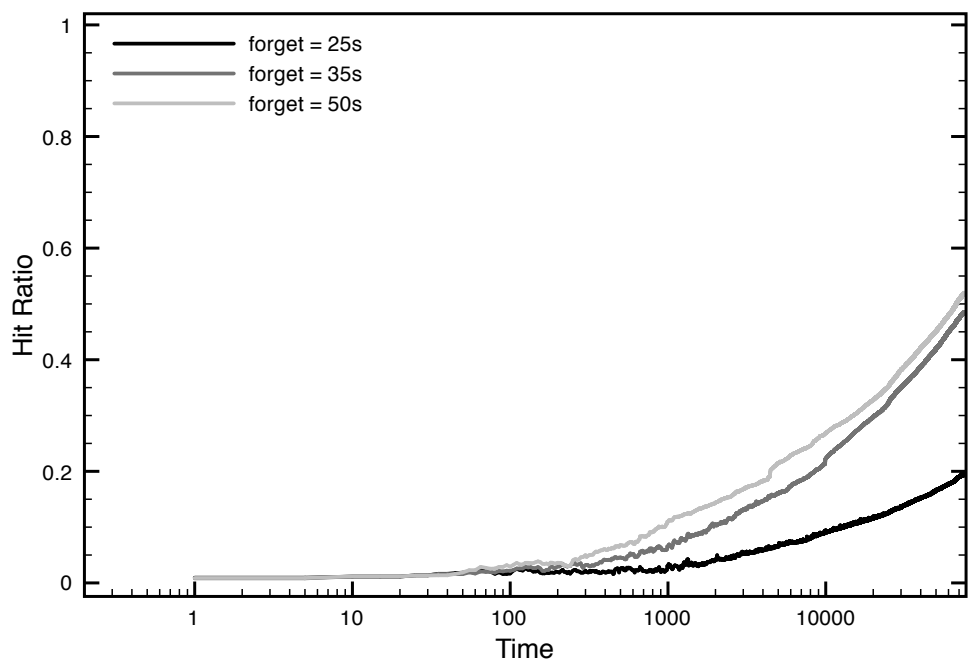

(b) Dest. Comm.

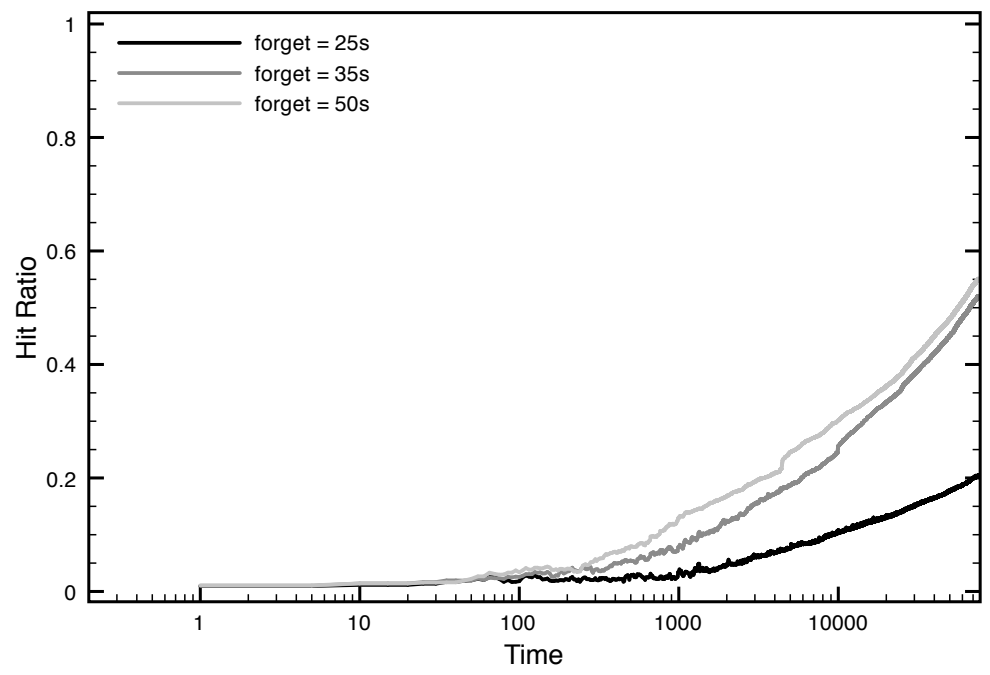

(c) Ext. Comm.

Fig. 15. Hit Ratio of a tagged traveller w.r.t its home (a), destination (b) and external (c) communities, for different values of the *forget* threshold

## Acknowledgment

This work is supported by the EINS (FP7-FIRE 288021) EC project.

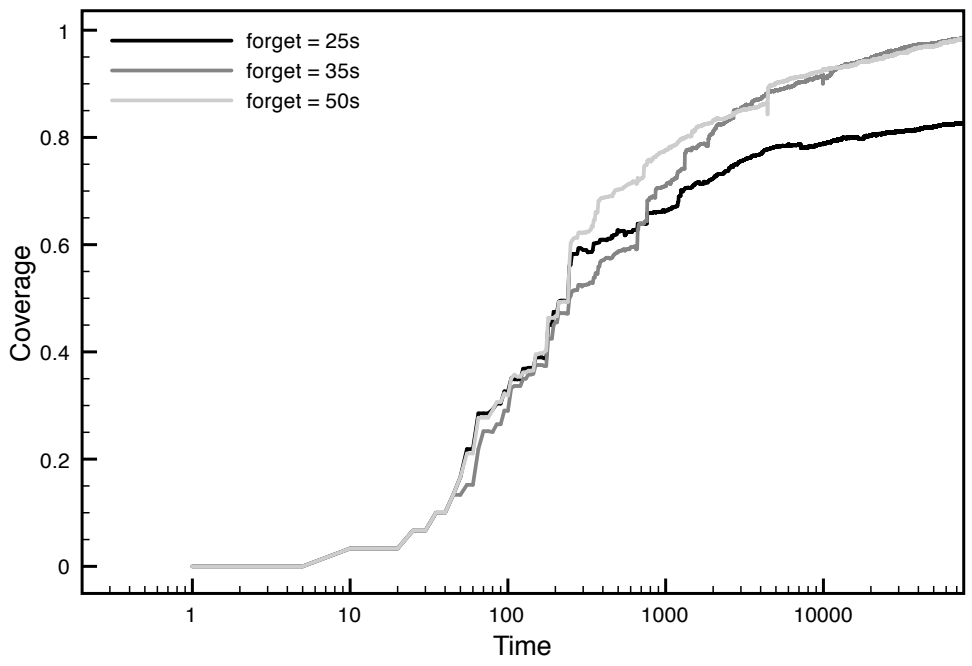

(a) Home Comm.

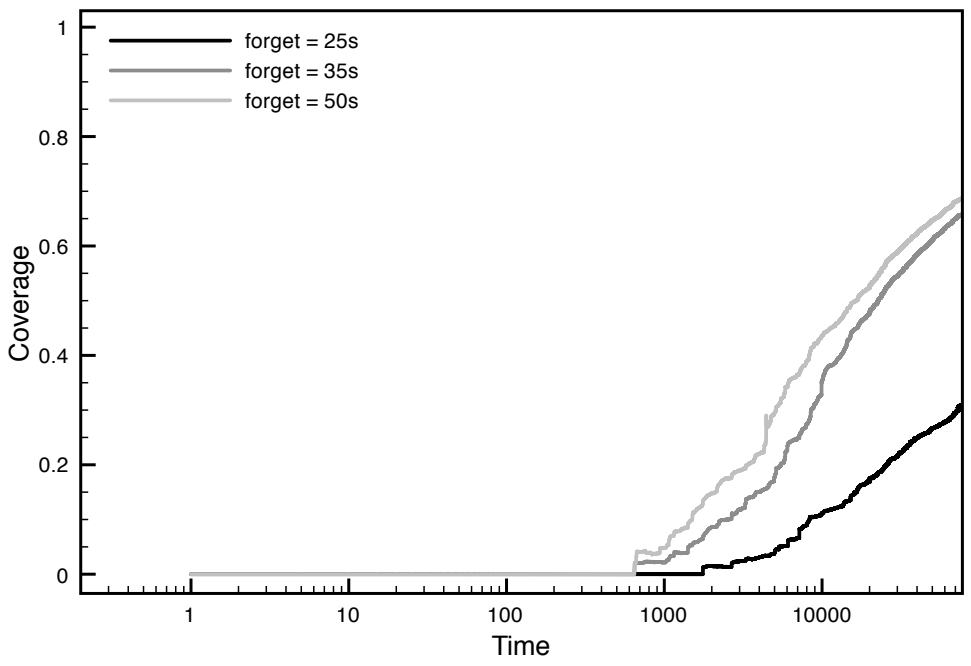

(b) Dest. Comm.

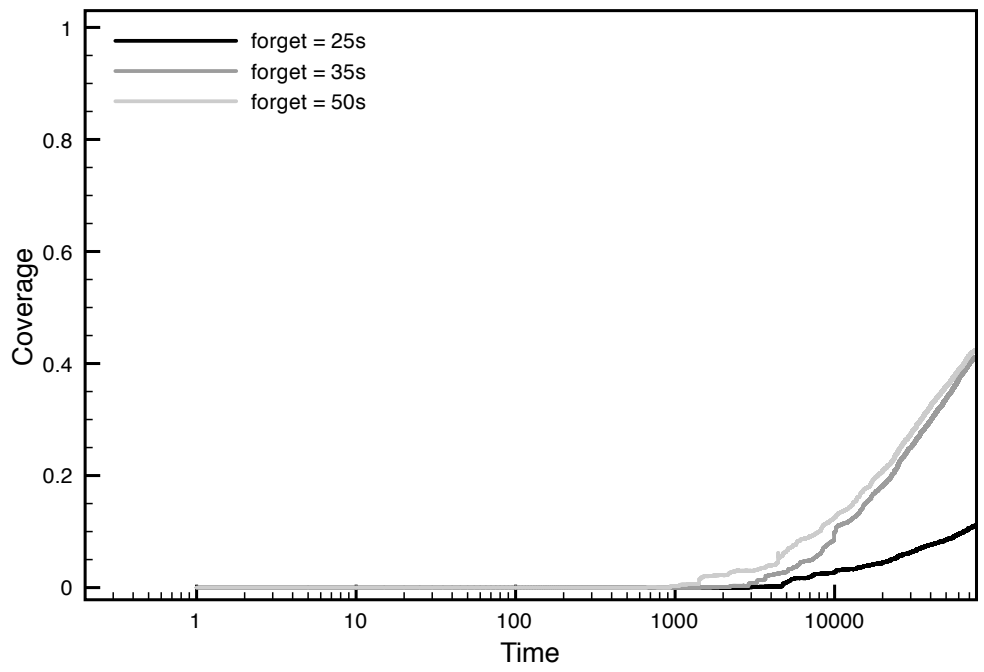

(c) Ext. Comm.

Fig. 16. Coverage of a tagged traveller w.r.t its home (a), destination (b) and external (c) communities, for different values of the *forget* threshold

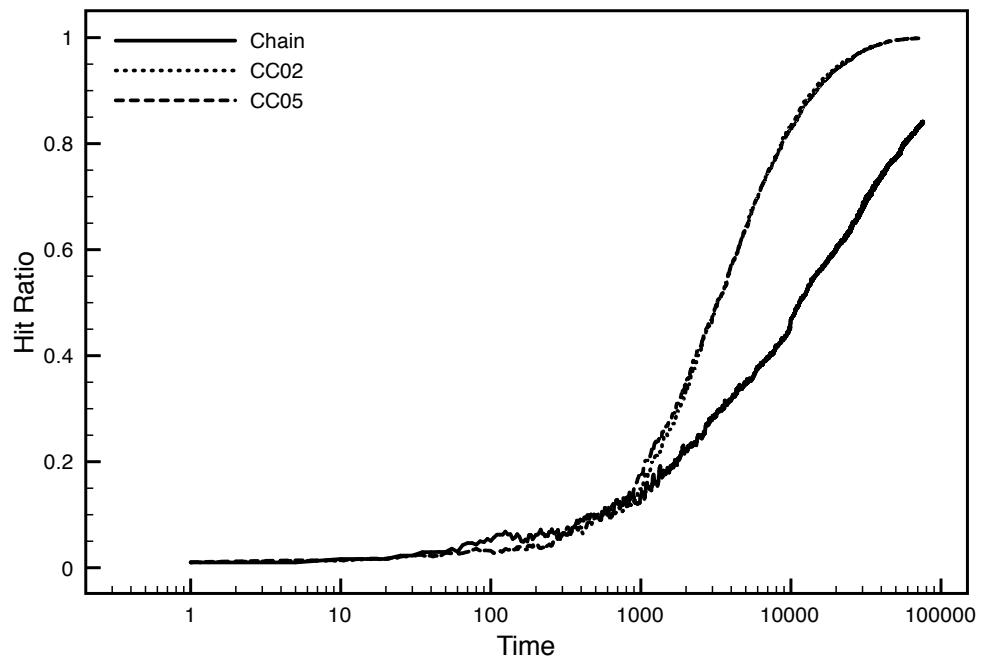

(a) Home Comm.

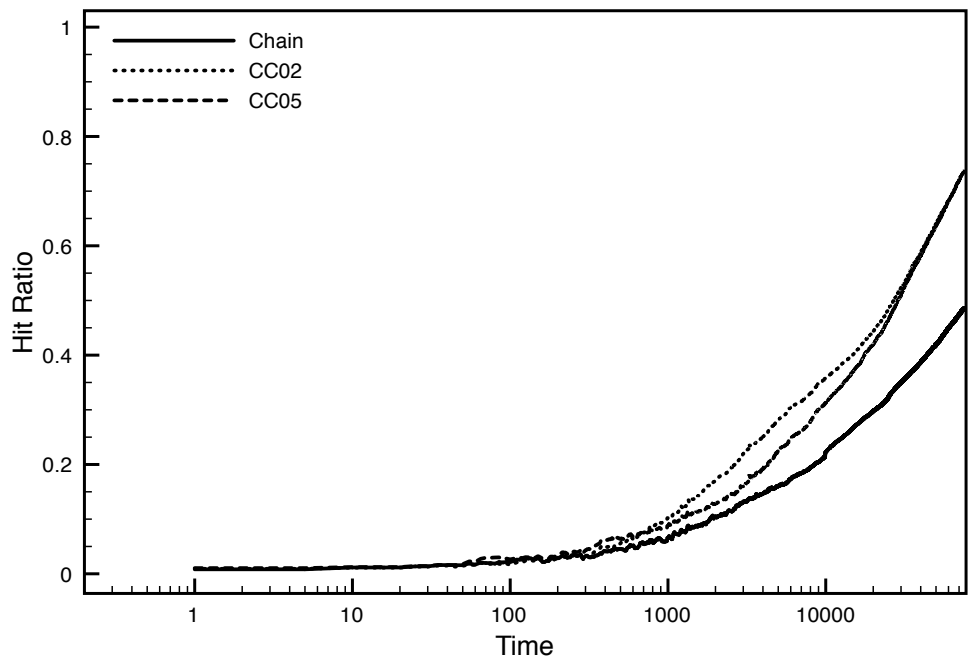

(b) Dest. Comm.

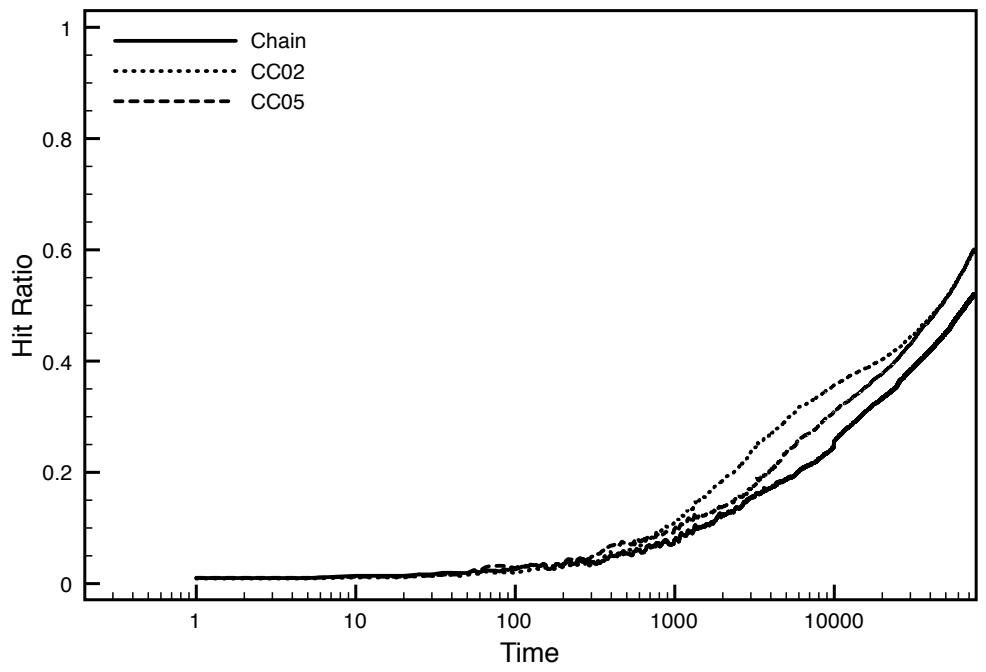

(c) Ext. Comm.

Fig. 17. Hit Ratio of a tagged traveller w.r.t its home (a), destination (b) and external (c) communities, for different organizations of the Locations SAN

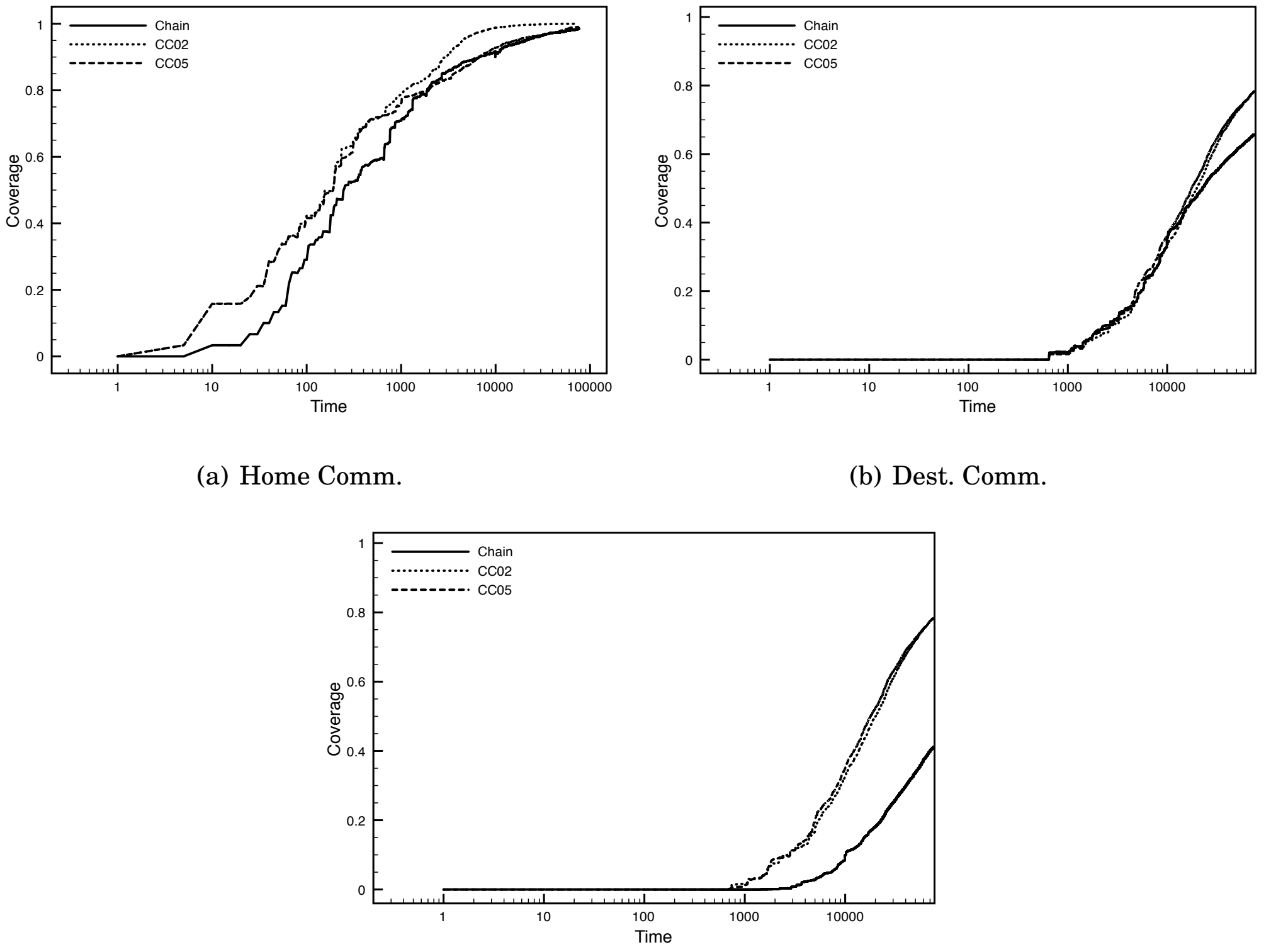

(a) Home Comm.

(b) Dest. Comm.

(c) Ext. Comm.

Fig. 18. Coverage of a tagged traveller w.r.t its home (a), destination (b) and external (c) communities, for different organizations of the Locations SAN